\documentclass[a4paper]{article}

\usepackage[margin=1in]{geometry} % full-width

\usepackage{amsmath}
\usepackage{amsthm}
\usepackage{amssymb}

\usepackage[utf8]{inputenc}
\usepackage[hidelinks]{hyperref}
\hypersetup{
	unicode,
	pdfauthor={Author One, Author Two, Author Three},
	pdftitle={AGMO2025},
	pdfsubject={A simple article template},
	pdfkeywords={article, template, simple},
	pdfproducer={LaTeX},
	pdfcreator={pdflatex}
}

\usepackage[sort&compress,numbers,square]{natbib}
\theoremstyle{plain}

\theoremstyle{definition}

\usepackage{graphicx, color}
\graphicspath{{fig/}}

\usepackage{comment}
\usepackage[dvipsnames,svgnames]{xcolor}

\newcommand\blfootnote[1]{%
  \begingroup
  \renewcommand\thefootnote{}\footnote{#1}%
  \addtocounter{footnote}{-1}%
  \endgroup
}

\def\be{\begin{equation}}
	\def\ee{\end{equation}}
\def\bse{\begin{subequations}}
	\def\ese{\end{subequations}}
\usepackage{algorithm, algpseudocode} 
\usepackage{mathrsfs} 

\usepackage{lipsum}
\numberwithin{equation}{section}
\title{On the integrable six-wave interaction system\\and its space-time shifted reduction}
\author{Mark J. Ablowitz$^1$,\;\; Ramesh Gupta$^2$,\;\; Ziad H. Musslimani$^2$,\;\; Nicholas J. Ossi$^3$}
\date{
	\small{$^1$ Department of Applied Mathematics, University of Colorado, Boulder, CO 80309\\\vspace{.1cm}
	$^2$ Department of Mathematics, Florida State University, Tallahassee, FL 32306\\\vspace{.1cm}
    $^3$ Department of Mathematics, State University of New York, Buffalo, NY 14260}
}
\begin{document}
	\maketitle
%%%%%%%%%%%%%%%%%%%%%%%%%%%%%%%%%%%%%%%%%%%%%%%%%%%%%%%%%%%%%%%%%%%%%
	\begin{abstract}
%%%%%%%%%%%%%%%%%%%%%%%%%%%%%%%%%%%%%%%%%%%%%%%%%%%%%%%%%%%%%%%%%%%%%	
The multi-dimensional  six-wave interaction system is derived in the context of nonlinear optics. Starting from Maxwell's equations, 
a reduced system of equations governing the dynamics of the electric and polarization fields are obtained. Using a space-time multi-scale 
asymptotic expansion, a hierarchy of coupled equations describing the spatio-temporal evolution of the perturbed electric and polarization fields 
are derived. The leading order equation admits a six-wave ansatz  satisfying a triad resonance condition.
By removing secular terms at next order, a first order in space and time quadratically nonlinear coupled six-wave interaction system is obtained. 
This resulting system is tied to its integrable counterpart which was postulated by Ablowitz and Haberman in the 1970s.
A reduction to a space-time shifted nonlocal three-wave system is presented. The resulting system is solved using the inverse scattering transform,
which employs nonlocal symmetries between the associated eigenfunctions and scattering data; soliton solutions are then obtained. Finally, an infinite set of conservation laws for the six-wave system is derived; one is shown to be connected to its Hamiltonian structure. 
%%%%%%%%%%%%%%%%%%%%%%%%%%%%%%%%%%%%%%%%%%%%%%%%%%%%%%%%%%%%%%%%%%
	\end{abstract}
%%%%%%%%%%%%%%%%%%%%%%%%%%%%%%%%%%%%%%%%%%%%

\blfootnote{We are honored to contribute this article to this special issue in memory of Professor Vladimir E. Zakharov. Vladimir Zakharov was a giant in the field of nonlinear waves. He made pioneering and  landmark contributions to the study of wave turbulence, integrable systems, water waves, plasma physics amongst other fields of study. His work will live forever.}

%%%%%%%%%%%%%%%%%%%%%%%%%
	\tableofcontents
%%%%%%%%%%%%%%%%%%%%%%%%%%%%%%%%%%%%%%%%%%%%%%%%%%%%%%%%%%%%%%%%%%%%%
\section{Introduction}
Multi-wave resonant interactions constitute one of the fundamental concepts ubiquitous to nonlinear dispersive systems. In its simplest form, 
it amounts to a nonlinear conversion process by which two incoming waves corresponding to wave vectors ${\bf k}_1$ and ${\bf k}_2$ and frequencies 
$\omega_1\equiv \omega({\bf k}_1)$ and 
$\omega_2 \equiv \omega({\bf k}_2)$, where $\omega({\bf k})$ is the dispersion relation, combine to generate a new (third) wave with
frequency $\omega_3$ and wave vector ${\bf k}_3$ satisfying the so-called resonant triad: 
%%%%%%%%%%%%%%%%%%%%%%%%%%%%%%%%%
\begin{equation}
    \label{resonance}
    \mathbf{k}_{1}+\mathbf{k}_{2}+\mathbf{k}_{3}=0\,,\;\;\;\;\omega_{1}+\omega_{2}+\omega_{3}=0\,,
\end{equation}
%%%%%%%%%%%%%%%%%%%%%%%%%%%%%%%%%%%%%%%%%%%%%%%
or three-wave resonant interaction. When ${\bf k}_2={\bf k}_1$ we term this a two-wave resonant interaction. Physically speaking, such coherent ``nonlinear addition" of waves are special: it requires a match between {\it both} the wave vectors and their respective frequencies -- a property not every dispersive medium allows. For example, in deep water without surface tension the dispersion relation $\omega = \pm \sqrt{g|{\bf k}|}$ does not satisfy the resonance condition (\ref{resonance}). In that case a four-wave resonant interaction is found.
In fluid mechanics/ocean dynamics, Phillips \cite{phillips1960,phillips1966}
showed that  multi-wave wave resonant interactions are important; subsequently, in the context of deep water waves, Benny \cite{benney1962}  found temporal slowly varying four-wave  interaction equations  that satisfied a four wave resonance condition. McGoldrich \cite{mcgoldrich1965} showed that three-wave or triad resonance satisfying conditions such as (\ref{resonance}) can occur in deep water waves with surface tension. Simmons \cite{simmons1969} then gave a geometrical argument for three-wave resonance in deep water waves with surface tension  and found an asymptotic reduction  to three wave equations with slowly varying amplitudes in both space and time. Water wave experiments were carried out that demonstrated the interaction phenomena by McGoldrick \cite{McGoldrick 1970}, and Henderson and Hammack \cite{HendersonHammack 1987,HammackHenderson 1993}. A general form of the interaction equations in conservative systems was studied by
Hasselmann \cite{Hasselman1967}; see also papers by Zakharov and co-authors: \cite{DyachZakKuz1996},\cite{DyachKorotZak2003}. 

Recently Ablowitz, Luo, and Musslimani \cite{ALM_six} showed that a model for finite depth water waves with surface tension has an asymptotic reduction to an integrable  six-wave interaction system in both $1+1$ and $2+1$ dimensions found by Ablowitz and Haberman \cite{AblHab1975} in the 1970s. Detailed discussion of the integrable nature and solutions of the three wave system in $1+1$ dimensions was carried out by Zakharov and Manakov  \cite{ZakMan1975} and Kaup \cite{Kaup1976} and in $2+1$ dimensions by Kaup \cite{Kaup1981} and Fokas and Ablowitz \cite{FoAbl1983}.

In nonlinear optics, such resonant wave mixing was demonstrated in quadratic media where second harmonic generation as well as parametric three and four wave mixing processes have been {analyzed and observed; see e.g. \cite{AkhKhok1972,Agrawal3rdEd2001,AS_Book}. In this article we show that the $2+1$ dimensional six-wave system \cite{AblHab1975} can be derived in nonlinear optics; we do this two different ways: i) via a susceptibility model -- see the system (\ref{6wWaveEqA}) and ii) via  a laser model -- see the system (\ref{Laser1})-(\ref{Laser1}).
~This system is equivalent to the integrable system (\ref{six_wave_QR}) which has reductions (\ref{reduc_classical}) -(\ref{three_wave_shifted}). 
The system (\ref{six_wave_QR}) corresponds, in general, to a complex function reduction of the physical system. Complex reductions/solutions of physical systems have been intensively studied for many years;  see e.g. self-dual reductions of Yang–Mills and Einstein equations \cite{Ward1977,Gibbons1990,AblCl1991} and water waves  \cite{LushKorot2016,Lush2016,Dyach2021}.

The six-wave system in both $1+1$ and $2+1$ dimensions are shown to be  Hamiltonian systems. Using the results from the integrability of the $1+1$ system \cite{ALM_three}  and space time reductions of NLS equations  \cite{AM_shifted}, \cite{AMO_shifted}, we find soliton solutions of the space-time shifted three wave systems. We compare these results with solitons of  the classical three wave system.

%%%%%%%%%%%%%%%%%%%%%%%%%%%%%%%%%%%%%%%%%%%%%%%%%%%%%
%%%%%%%%%%%%%%%%%%%%%%%%%%%%%%%%%%%%%%%%%%%%%%%%%%%%%%%%%%%%%%%%%%%%%
\section{The six-wave interaction system and reductions}
%%%%%%%%%%%%%%%%%%%%%%%%%%%%%%%%%%%%%%%%%%%%%%%%%%%%%%%%%%%%%%%%%%%%%
We begin with the above mentioned six-wave interaction system in the generic form for a $3\times3$ off-diagonal matrix system whose components satisfy
%%%%%%%%%%%%%%%%%%%%%%%%%%%%%%%%%
\begin{equation}
\label{six_wave_N_2}
    \partial_{t}N_{\ell j}(x,y,t) - \alpha_{\ell j}\partial_{x}N_{\ell j}(x,y,t) - \beta_{\ell j}\partial_{y}N_{\ell j}(x,y,t)
    =
    (\alpha_{\ell m}-\alpha_{mj})N_{\ell m}(x,y,t)N_{mj}(x,y,t)\,,
\end{equation}
%%%%%%%%%%%%%%%%%%%%%%%%%%%%%%%%%
where the indices $\ell, j, m\in1,2,3$ are distinct, i.e. $\ell\neq j,\;j\neq m,\;m\neq\ell$. Furthermore,  the constant off-diagonal matrices $\alpha , \beta$ are defined by: $\alpha_{23}=\alpha_{32}\equiv-C_{1}$, $\alpha_{31}=\alpha_{13}\equiv-C_{2}$, $\alpha_{12}=\alpha_{21}\equiv-C_{3}$; 
$\beta_{23}=\beta_{32}\equiv -C_1(C_2 + C_3)$, $\beta_{31}=\beta_{13}\equiv -C_{2}(C_1 + C_3)$, 
$\beta_{12}=\beta_{21}\equiv -C_{3} (C_1 + C_2).$ Here, we take
$C_{1}<C_{2}<C_{3}$. With the rescaling 
\bse
\label{scale_N_to_QR}
%%%%%%%%%%%%%%%%%%%%%%%%%%%%%%%%%
\begin{eqnarray}
\label{scale_N_to_Q}
    N_{23}=\frac{-iQ_{1}}{\sqrt{(C_{2}-C_{1})(C_{3}-C_{1})}}\,,\;N_{31}=\frac{-iQ_{2}}{\sqrt{(C_{2}-C_{1})(C_{3}-C_{2})}}\,,\;N_{12}=\frac{-iQ_{3}}{\sqrt{(C_{3}-C_{1})(C_{3}-C_{2})}}\,,\\
    N_{32}=\frac{-iR_{1}}{\sqrt{(C_{2}-C_{1})(C_{3}-C_{1})}}\,,\;N_{13}=\frac{iR_{2}}{\sqrt{(C_{2}-C_{1})(C_{3}-C_{2})}}\,,\;N_{21}=\frac{-iR_{3}}{\sqrt{(C_{3}-C_{1})(C_{3}-C_{2})}}\,,
\end{eqnarray}
%%%%%%%%%%%%%%%%%%%%%%%%%%%%%%%%%
\ese
(note that the lack of a negative sign in the relationship between $N_{13}$ and $R_{2}$ is purposeful) the system \eqref{six_wave_N} can be written as:
%%%%%%%%%%%%%%%%%%%%%%%%%%%%%%%%%
\bse
\label{six_wave_QR}
\begin{eqnarray}
%%%%%%%%%%%%%%%%%%%%%%%%%%%%%%%%%
\partial_{t}Q_{1}(x,y,t) + C^{(x)}_{1}\partial_{x}Q_{1}(x,y,t) + C^{(y)}_{1}\partial_{y}Q_{1}(x,y,t) &=& -iR_{2}(x,y,t)R_{3}(x,y,t)\,,\\
%%%%%%%%%%%%%%%%%%%%%%%%%%%%%%%%%
\partial_{t}Q_{2}(x,y,t)+C^{(x)}_{2}\partial_{x}Q_{2}(x,y,t) + C^{(y)}_{2}\partial_{y}Q_{2}(x,y,t)&=&-iR_{3}(x,y,t)R_{1}(x,y,t)\,,\\
%%%%%%%%%%%%%%%%%%%%%%%%%%%%%%%%%
\partial_{t}Q_{3}(x,y,t)+C^{(x)}_{3}\partial_{x}Q_{3}(x,y,t) + C^{(y)}_{3}\partial_{y}Q_{3}(x,y,t) &=&-iR_{1}(x,y,t)R_{2}(x,y,t)\,,\\
%%%%%%%%%%%%%%%%%%%%%%%%%%%%%%%%%
\partial_{t}R_{1}(x,y,t)+C^{(x)}_{1}\partial_{x}R_{1}(x,y,t) + C^{(y)}_{1}\partial_{y}R_{1}(x,y,t) &=&iQ_{2}(x,y,t)Q_{3}(x,y,t)\,,\\
%%%%%%%%%%%%%%%%%%%%%%%%%%%%%%%%%
\partial_{t}R_{2}(x,y,t)+C^{(x)}_{2}\partial_{x}R_{2}(x,y,t) + C^{(y)}_{2}\partial_{y}R_{2}(x,y,t) &=&iQ_{3}(x,y,t)Q_{1}(x,y,t)\,,\\
%%%%%%%%%%%%%%%%%%%%%%%%%%%%%%%%%
\partial_{t}R_{3}(x,y,t)+C^{(x)}_{3}\partial_{x}R_{3}(x,y,t) + C^{(y)}_{3}\partial_{y}R_{3}(x,y,t) &=&iQ_{1}(x,y,t)Q_{2}(x,y,t)\,.
%%%%%%%%%%%%%%%%%%%%%%%%%%%%%%%%%
\end{eqnarray}
\ese
%%%%%%%%%%%%%%%%%%%%%%%%%%%%%%%%%
Here, $C^{(x)}_{j} = C_{j}$ and $C^{(y)}_{1} = C_{1}(C_2+C_3),\, 
C^{(y)}_{2} = C_{2}(C_1+C_3),\, C^{(y)}_{3} = C_{3}(C_1+C_2)$; note also that $C^{(y)}_{1}<C^{(y)}_{2}<C^{(y)}_{3}$. 
%%%%%%%%%%%%%%%%%%%%%%%%%%%%%%%%%%%%%%%%%%%%%%%%%%%%%%%%%
Note that the system (2.3) is PT symmetric, i.e., it is invariant under the transformation 
%%%%%%%%%%%%%%%%
\begin{equation}
\label{transformation}
\{ Q_j (x,y,t), R_j (x,y,t) \} \longrightarrow \{ Q^*_j (-x,-y,-t), R^*_j (-x,-y,-t) \} \;.
\end{equation}
%%%%%%%%%%%%%%
We remark that PT symmetric integrable systems were originally proposed in \cite{AM_1, AM_2,AM_3}.
%%%%%%%%%%%%%%%%%%%%%%%%%%%%%%%%%%
Upon imposing the symmetry reduction 
%%%%%%%%%%%%%%%%%%%%%%%%%%%%%%%%%
\begin{equation}
\label{reduc_classical}
R_{j}(x,y,t)=-\tilde{\epsilon}_{j}Q_{j}^{*}(x,y,t)\,,\;\;\;\; \tilde{\epsilon}_{1} \tilde{\epsilon}_{2} \tilde{\epsilon}_{3}=-1\,,
\end{equation}
%%%%%%%%%%%%%%%%%%%%%%%%%%%%%%%%%
where $\tilde{\epsilon}_{j}=\pm 1,\, j=1,2,3,$ the system \eqref{six_wave_QR} becomes the classical three-wave resonant interaction equations
%%%%%%%%%%%%%%%%%%%%%%%%%%%%%%%%%
\bse
\label{three_wave_classical}
\begin{eqnarray}
%%%%%%%%%%%%%%%%%%%%%%%%%%%%%%%%%%%%%%%%%%%%%%%%%%
    \partial_{t}Q_{1} + C^{(x)}_{1}\partial_{x}Q_{1}  + C^{(y)}_{1}\partial_{y}Q_{1}  &=&i\tilde{\epsilon}_{1}Q_{2}^{*} Q_{3}^{*}\,,\\
    %%%%%%%%%%%%%%%%%%%%%%%%%%%%%%%%%%%%%%%%%%%%%%%%%%
    \partial_{t}Q_{2}+C^{(x)}_{2}\partial_{x}Q_{2} + C^{(y)}_{2}\partial_{y}Q_{2} &=&i\tilde{\epsilon}_{2}Q_{3}^{*} Q_{1}^{*} \,,\\
    %%%%%%%%%%%%%%%%%%%%%%%%%%%%%%%%%%%%%%%%%%%%%%%%%%
    \partial_{t}Q_{3}+C^{(x)}_{3}\partial_{x}Q_{3} + C^{(y)}_{3}\partial_{y}Q_{3} &=&i\tilde{\epsilon}_{3}Q_{1}^{*} Q_{2}^{*} \,.
    %%%%%%%%%%%%%%%%%%%%%%%%%%%%%%%%%%%%%%%%%%%%%%%%%%
\end{eqnarray}
\ese
%%%%%%%%%%%%%%%%%%%%%%%%%%%%%%%%%
Alternatively, a reverse space-time nonlocal symmetry reduction was introduced in \cite{ALM_three}. Presently, we generalize this reduction, as was done in \cite{AM_shifted,AMO_shifted} for NLS-type systems, to include nonlocal space-time shifts in addition to reflections. In particular, if we enforce
%%%%%%%%%%%%%%%%%%%%%%%%%%%%%%%%%
\begin{eqnarray}
\label{reduc_shifted}
    R_{j}(x,y,t)=(-1)^{j+1}\epsilon_{j}Q_{j}^{*}(x_{0}-x,y_{0}-y, t_{0}-t)\,,\;\;\;\;\epsilon_{1}\epsilon_{2}\epsilon_{3}=+1\,,
\end{eqnarray}
%%%%%%%%%%%%%%%%%%%%%%%%%%%%%%%%%
then \eqref{six_wave_QR} reduces to the complex space-time shifted three-wave interaction system:
\bse
\label{three_wave_shifted}
%%%%%%%%%%%%%%%%%%%%%%%%%%%%%%%%%
\begin{eqnarray}
%%%%%%%%%%%%%%%%%%%%%%%%%%%%%%%%%
     \partial_{t}Q_{1}+C^{(x)}_{1}\partial_{x}Q_{1} + C^{(y)}_{1}\partial_{y}Q_{1}
     &=&i\epsilon_{1}Q_{2}^{*}(x_{0}-x,y_{0}-y,t_{0}-t)Q_{3}^{*}(x_{0}-x,y_{0}-y,t_{0}-t)\,,\\
     %%%%%%%%%%%%%%%%%%%%%%%%%%%%%%%%%
     \partial_{t}Q_{2}+C^{(x)}_{2}\partial_{x}Q_2+ C^{(y)}_2\partial_{y}Q_2
    &=&-i\epsilon_{2}Q_{3}^{*}(x_{0}-x,y_{0}-y,t_{0}-t)Q_{1}^{*}(x_{0}-x,y_{0}-y,t_{0}-t)\,,\\
    %%%%%%%%%%%%%%%%%%%%%%%%%%%%%%%%%
    \partial_{t}Q_{3}+C^{(x)}_{3}\partial_{x}Q_3 + C^{(y)}_{3}\partial_{y}Q_3
    &=&i\epsilon_{3}Q_{1}^{*}(x_{0}-x,y_{0}-y,t_{0}-t)Q_{2}^{*}(x_{0}-x,y_{0}-y,t_{0}-t)\,.
    %%%%%%%%%%%%%%%%%%%%%%%%%%%%%%%%%
\end{eqnarray}
\ese
%%%%%%%%%%%%%%%%%%%%%%%%%%%%%%%%%
Note that this is not a translationally invariant system in space or time. However, the shifting parameters supply \eqref{three_wave_shifted} with a ``pseudo-translational symmetry" in the sense that translating a solution of \eqref{three_wave_shifted} gives a solution of another system in the same family. Specifically, if \eqref{three_wave_shifted} is solved by $Q_{j}(x,y,t)$, $j=1,2,3$, then 
%%%%%%%%%%%%%%%%%%%%%%%%%%%%%%%%%
\begin{equation}
\tilde{Q}_{j}(x,y,t)\equiv Q_{j}\Big(x+\frac{1}{2}(x_{0}-\tilde{x}_{0}) ,y+\frac{1}{2}(y_{0}-\tilde{y}_{0}), t+\frac{1}{2}(t_{0}-\tilde{t}_{0})\Big)\,,
\end{equation}
%%%%%%%%%%%%%%%%%%%%%%%%%%%%%%%%%
solves the same system but with $x_{0}, y_0$ and $t_{0}$ replaced by $\tilde{x}_{0}, \tilde{y}_{0}$ and $\tilde{t}_{0}$, respectively. 
%%%%%%%%%%%%%%%%%%%%%%%%%%%%%%%%%%%%%%%%%%%%%%%%%%%%%%%%%%%%%%%%%%
\section{Derivation of the six-wave system in nonlinear optics}
%%%%%%%%%%%%%%%%%%%%%%%%%%%%%%%%%%%%%%%%%%%%%%%%%%%%%%%%%%%%%%%%%%
Before discussing the integrability of \eqref{three_wave_shifted}, it is worth reiterating that the general six-wave system \eqref{six_wave_QR} has been shown to arise as a complex asymptotic reduction of dispersive models with quadratic nonlinearity. In \cite{ALM_six}, the six-wave system was obtained from the classical water-wave equations with surface tension. Here, we demonstrate that this system can be derived in the context of optical media with quadratically nonlinear polarization. 
%%%%%%%%%%%%%%%%%%%%%%%%%%%%%%%%%%%%%%%%%%%%%%%%%%%%%%%%%%%%%%%%%
\subsection{Susceptibility model}
%%%%%%%%%%%%%%%%%%%%%%%%%%%%%%%%%%%%%%%%%%%%%%%%%%%%%%%%%%%%%%%%
We start by considering Maxwell's equations that govern the dynamics of the electric and magnetic fields,
$\mathbf{E}$ and $\mathbf{H}$, respectively
%%%%%%%%%%%%%%%%%%%%%%%%%%%%%%%%%
\begin{equation}
\label{maxwell-eqns}
    \nabla\cdot\mathbf{D}=0\,,\qquad\nabla\cdot\mathbf{B}=0\,, \qquad
    %%%%%%%%%%%%%%%%%%%%%%%%%%%%%%%%%
\frac{ \partial \mathbf{D}}{\partial t} = \nabla\times\mathbf{H}\,, \qquad
    %%%%%%%%%%%%%%%%%%%%%%%%%%%%%%%%%
 \frac{\partial \mathbf{B}}{\partial t}  =  -\nabla\times\mathbf{E} \,.
\end{equation}
%%%%%%%%%%%%%%%%%%%%%%%%%%%%%%%%%
In (\ref{maxwell-eqns}),  $\mathbf{B} \equiv \mu_{0}\mathbf{H}$ is the magnetic induction while $\mathbf{D} \equiv \varepsilon_{0} (\mathbf{E}+\mathbf{P(\mathbf{E})})$ is the electric displacement with $\mathbf{P}$ being the polarization vector (which depends on the electric field). 
The constants $\mu_{0}$ and $\varepsilon_{0}$ denote the magnetic permeability and the electric permittivity of free space respectively. After some algebra, Eqns.~(\ref{maxwell-eqns}) give rise to the coupled system:
%%%%%%%%%%%%%%%%%%%%%%%%%%%%%%%%%
\begin{align}
    \label{Maxwell_vector-1}
   &  \frac{1}{c^2} \partial^2_t  \left( \mathbf{E} + \mathbf{P} (\mathbf{E}) \right) 
   + \nabla(\nabla\cdot\mathbf{E}) - \nabla^{2}\mathbf{E}=0 \;,
   \\
  \label{Maxwell_vector-2}
  & \nabla \cdot \left( \mathbf{E} + \mathbf{P} (\mathbf{E})  \right) = 0 \;,
\end{align}
%%%%%%%%%%%%%%%%%%%%%%%%%%%%%%%%%
where $c^{2}=1/\mu_{0}\varepsilon_{0}$. Each vector field depends on the spatial variables $\mathbf{r}=(x,y,z)$ and time $t$. Here, $\nabla^2$ denotes the three-dimensional spatial Laplacian. The polarization field has two contributions: linear and nonlinear. Depending on the material response, the nonlinear part could depend quadratically or cubically on the electric field. Since in this paper we are interested in deriving six-wave interaction equations, we will assume that the medium response is quadratic (and ignore higher order contributions). With this in mind, the polarization vector is written in the form:
%%%%%%%%%%%%%%%%%%%%%%%%%%%%%%%%%
\begin{equation}
    \label{polarization}
   \mathbf{P} = \underbrace{\chi^{(L)} \star \mathbf{E}}_{\mathbf{P}^{(L)} } 
   + \underbrace{\chi^{(NL)} \star \mathbf{E} \mathbf{E} }_{\mathbf{P}^{(NL)} } +\cdots \;,
\end{equation}
%%%%%%%%%%%%%%%%%%%%%%%%%%%%%%%%%
where $\chi^{(L)}$ and $\chi^{(NL)}$ are rank 2 and 3 tensors that depend only on time (both are set to zero for $t<0$) and star denotes temporal convolution. Specifically, we have
%%%%%%%%%%%%%%%%%%%%%%%%%%%%%%%%%
\begin{equation}
    \label{polarization-L}
   P_j^{(L)} (\mathbf{r} ,t)= \sum_{k=1}^3 \int_{-\infty}^{+\infty} \chi_{jk}^{(L)}(t-t_1) E_k (\mathbf{r}, t_1) dt_1 \;,
\end{equation}
%%%%%%%%%%%%%%%%%%%%%%%%%%%%%%%%%
%%%%%%%%%%%%%%%%%%%%%%%%%%%%%%%%%
\begin{equation}
    \label{polarization-NL}
   P_j^{(NL)} (\mathbf{r} ,t) = \sum_{k,\ell =1}^3 \int_{-\infty}^{+\infty} \chi_{jk\ell}^{(NL)}(t-t_1, t-t_2) 
   E_k (\mathbf{r}, t_1)E_\ell (\mathbf{r} ,t_2) dt_1 dt_2 \;,
\end{equation}
%%%%%%%%%%%%%%%%%%%%%%%%%%%%%%%%%
where the indices $j=1,2,3$ correspond to the $x,y,z$ directions respectively. Consider the asymptotic expansion
%%%%%%%%%%%%%%%%%%%%%%%%%%%%%%
\begin{equation}
    \label{E-field-asym}
   \mathbf{E} = \epsilon  \mathbf{E}^{(1)} + \epsilon^2  \mathbf{E}^{(2)} + \cdots\;.
   \end{equation}
%%%%%%%%%%%%%%%%%%%%%%%%%%%%%%%%%
Substituting (\ref{E-field-asym}) into (\ref{Maxwell_vector-1}) and (\ref{Maxwell_vector-2}) leads to
%%%%%%%%%%%%%%%%%%%%%%%%%%%%%%%%%%%%%%%%%%%%%%%%%%%%%%%%%%%%%%%%%%
%%%%%%%%%%%%%%%%%%%%%%%%%%%%%%%%%%%%%%%%%%%%%%%%%%%%%%%%%%%%%%%%%%
\begin{align}
    \label{Maxwell_vector-1-asym-2}
   &  \frac{1}{c^2} \partial^2_t  \left(   \mathbf{E}^{(1)} +  \mathbf{P}^{(L)} (\mathbf{E}^{(1)}) + \epsilon  \mathbf{E}^{(2)}  
    + \epsilon \mathbf{P}^{(L)} (\mathbf{E}^{(2)}) + \epsilon \mathbf{P}^{(NL)} (\mathbf{E}^{(1)})  \right) 
    \nonumber \\
  & +  \nabla(\nabla\cdot\mathbf{E}^{(1)}) -   \nabla^{2}\mathbf{E}^{(1)}  + \epsilon \nabla(\nabla\cdot\mathbf{E}^{(2)}) 
  -  \epsilon \nabla^{2}\mathbf{E}^{(2)} 
  =
   O(\epsilon^2) \;,
 \end{align}
 %%%%%%%%%%%%%%%%%%%%%%%%%%%%%%%%%%%%%%%%%%%%%%%%%%%%%%%%%%%%%%%%
 and
 %%%%%%%%%%%%%%%%%%%%%%%%%%%%%%%%%%%%%%%%%%%%%%%%%%%%%%%%%%%%%%%%%%
\begin{equation}
  \label{Maxwell_vector-2-asym-2}
  \nabla \cdot \left(   \mathbf{E}^{(1)} +   \mathbf{P}^{(L)} (\mathbf{E}^{(1)}) + \epsilon \mathbf{E}^{(2)} 
   + \epsilon \mathbf{P}^{(L)} (\mathbf{E}^{(2)}) + \epsilon \mathbf{P}^{(NL)} (\mathbf{E}^{(1)}) \right) 
  = 
   O(\epsilon^2) \;.
\end{equation}
%%%%%%%%%%%%%%%%%%%%%%%%%%%%%%%%%
Next, introduce the slow time and space variables 
%%%%%%%%%%%%%%%%%%%%%%%%%%%%%%%%%
\begin{equation}
\label{perturb}
     T = \epsilon t\,,\;\;\;\ \mathbf{R} =\epsilon \mathbf{r}\,,\;\;\;
     \nabla \rightarrow \nabla_\mathbf{r} + \epsilon \nabla_{\mathbf{R}} \;,\;\;\;\;\; \partial_t \rightarrow \partial_t + \epsilon \partial_T \;.
     \end{equation}
%%%%%%%%%%%%%%%%%%%%%%%%%%%%%%%%%
Substituting (\ref{perturb}) into (\ref{Maxwell_vector-1-asym-2}) and (\ref{Maxwell_vector-2-asym-2}) leads to
%%%%%%%%%%%%%%%%%%%%%%%%%%%%%%%%%%%%%%%%%%%%%%%%%%%%%%%%%%%%%%%%%%
\begin{align}
    \label{Maxwell_vector-1-asym-3}
   &  \frac{1}{c^2}\left(  \partial^2_t + 2\epsilon \partial_{tT}  \right) \left(   \mathbf{E}^{(1)} +  \mathbf{P}^{(L)} (\mathbf{E}^{(1)}) + \epsilon  \mathbf{E}^{(2)}  
    + \epsilon \mathbf{P}^{(L)} (\mathbf{E}^{(2)}) + \epsilon \mathbf{P}^{(NL)} (\mathbf{E}^{(1)})  \right) 
    %%%%%%%%%%%%%%%%%%%%%%%%%%%%%%%
    \nonumber \\
  & +  \nabla_\mathbf{r}(\nabla_\mathbf{r}\cdot\mathbf{E}^{(1)}) 
  + \epsilon  \nabla_\mathbf{r}(\nabla_\mathbf{R}\cdot\mathbf{E}^{(1)}) 
  + \epsilon \nabla_\mathbf{R}(\nabla_\mathbf{r}\cdot\mathbf{E}^{(1)})
  %%%%%%%%%%%%%%%%%%%%%%%%%%%%%%%
    \nonumber \\
  & 
  -   \nabla_\mathbf{r}^{2} \mathbf{E}^{(1)} - 2\epsilon \left(\nabla_\mathbf{R} \cdot \nabla_\mathbf{r}\right) \mathbf{E}^{(1)} 
   + \epsilon \nabla_\mathbf{r}(\nabla_\mathbf{r}\cdot\mathbf{E}^{(2)}) 
  -  \epsilon \nabla_\mathbf{r}^{2}\mathbf{E}^{(2)} 
  %%%%%%%%%%%%%%%%%%%%%%%%%%%%%%
  =
   O(\epsilon^3) \;,
 \end{align}
 %%%%%%%%%%%%%%%%%%%%%%%%%%%%%%%%%%%%%%%%%%%%%%%%%%%%%%%%%%%%%%%%%%
   and
   %%%%%%%%%%%%%%%%%%%%%%%%%%%%%%%%%%%%%%%%%%%%%%%%%%%%%%%%%%%%%%%%%
   \begin{equation}
  \label{Maxwell_vector-2-asym-3}
  \left( \nabla_\mathbf{r} + \epsilon \nabla_{\mathbf{R}} \right)\cdot \left(   \mathbf{E}^{(1)} +   \mathbf{P}^{(L)} (\mathbf{E}^{(1)}) + \epsilon \mathbf{E}^{(2)} 
   + \epsilon \mathbf{P}^{(L)} (\mathbf{E}^{(2)}) + \epsilon \mathbf{P}^{(NL)} (\mathbf{E}^{(1)}) \right) 
  = 
   O(\epsilon^3) \;.
\end{equation}
%%%%%%%%%%%%%%%%%%%%%%%%%%%%%%%%%
At this stage, we still need to determine if the polarization fields, $\mathbf{P}^{(L)}$ and $\mathbf{P}^{(NL)}$ have any ``hidden" dependence 
on $\epsilon$ due to the action of the space-time multi-scales on their convolutional representations. Before resolving this issue, we first make the following assumptions: The susceptibility tensor $\chi_{jk}^{(L)}$ vanishes whenever $j\ne k.$ Furthermore, the only nonzero elements of the $\chi_{jk\ell}^{(NL)}$ 
tensor arise from indices satisfying $j=k=\ell$ and  assume that the leading order electric field
is in the $\hat{\mathbf{z}}$ direction and is $z$ independent. Thus, we write 
%%%%%%%%%%%%%%%%%%%%%%%%%%%%%%%%%
\begin{equation}
 \mathbf{E}^{(1)} \left( \mathbf{r},t; \mathbf{R} , T\right) = E^{(1)}_z  \left( \mathbf{r}_{\perp},t; \mathbf{R}_{\perp} , T\right) \hat{\mathbf{z}} \;,
\end{equation}
%%%%%%%%%%%%%%%%%%%%%%%%%%%%%%%%%
(recall: $E^{(1)}_z=E^{(1)}_1$) where $\mathbf{r}_{\perp}$ is the two-dimensional transverse $(x,y)$ plane and $\mathbf{R}_{\perp} \equiv \epsilon \mathbf{r}_{\perp}.$ Note that subscripts appearing on the electric and polarization fields do {\it not} represent partial derivatives. We next assume that the leading order electric field takes the form 
%%%%%%%%%%%%%%%%%%%%%%%%%%%%%%%%%
\begin{equation}
\label{six_wave_ansatz_E}
    E^{(1)}_z  \left( \mathbf{r}_{\perp},t; \mathbf{R}_{\perp} , T\right) 
    =
    \sum_{j=1}^{3}\Big[ A_{j}(\mathbf{R}_{\perp} , T) e^{i\theta_{j}( \mathbf{r}_{\perp},t ) }
    + B_{j}(\mathbf{R}_{\perp} , T) e^{-i\theta_{j}( \mathbf{r}_{\perp},t ) }\Big]\,,
\end{equation}
%%%%%%%%%%%%%%%%%%%%%%%%%%%%%%%%%
where $\theta_j \equiv \mathbf{k}_j\cdot\mathbf{r}_{\perp} - \omega_j t$ and $\omega_j  \equiv \omega ( \mathbf{k}_j).$ The wave vectors  $\mathbf{k}_j$
and the corresponding frequencies $\omega_j$  are assumed to satisfy the resonant triad condition for all $\mathbf{r}_{\perp},t$:
%%%%%%%%%%%%%%%%%%%%%
\begin{equation}
\label{theta-resonance-triad}
 \theta_{1}+\theta_{2}+\theta_{3}=0\,.
\end{equation}
%%%%%%%%%%%%%%%%%%%%%%%
The dispersion relation $\omega (\mathbf{k})$ will be later derived for a generic susceptibility tensor. A typical case will be given for which an explicit formula for the dispersion relation is readily available and can be used to establish a resonant triad. Note that in (\ref{six_wave_ansatz_E}),  
we do not assume $A_{j}$ and $B_{j}$ are complex conjugates of each other. This is a key assumption that enables us to derive the triad six-wave interaction equations. If instead, the relation $A_{j}=B^*_{j}$ is imposed (at this stage) 
this would lead to the three wave triad system. 

Next, we examine the multi-scale structure of $\mathbf{P}^{(L)} (\mathbf{E}^{(1)}).$ From the assumption on the linear susceptibility $ \chi_{jk}^{(L)}$, and from (\ref{polarization-L}), we conclude that the leading order linear polarization field also
points in the $\hat{\mathbf{z}}$ direction and is $z$ independent. In this case, we have
%%%%%%%%%%%%%%%%%%%%%%%%%%%%%%%%%
\begin{equation}
 \mathbf{P}^{(L)} (\mathbf{E}^{(1)}) \left( \mathbf{r},t; \mathbf{R} , T\right) 
 = P^{(1)}_{L,z} (E_z^{(1)}) \left( \mathbf{r}_{\perp},t; \mathbf{R}_{\perp} , T\right) \hat{\mathbf{z}} \;,
\end{equation}
%%%%%%%%%%%%%%%%%%%%%%%%%%%%%%%%%
where, 
%%%%%%%%%%%%%%%%%%%%%%%%%%%%%%%%%
\begin{equation}
    \label{polarization-L-1}
    P^{(1)}_{L,z}  (E_z^{(1)}) \left( \mathbf{r}_{\perp},t; \mathbf{R}_{\perp} , T\right) 
   =  \int_{-\infty}^{+\infty} \chi_{zz}^{(L)}(t-t_1) E^{(1)}_z  \left( \mathbf{r}_{\perp},t_1; \mathbf{R}_{\perp} , \epsilon t_1\right)  dt_1 \;.
\end{equation}
%%%%%%%%%%%%%%%%%%%%%%%%%%%%%%%%%
It is customary in nonlinear optics to relabel $\chi_{33}^{(L)}$ and $\chi_{333}^{(NL)}$ as $\chi_{zz}^{(L)}$ and $\chi_{zzz}^{(NL)}$ (here, subscripts do not indicate partial derivatives). Substituting (\ref{six_wave_ansatz_E}) into (\ref{polarization-L-1}) leads to (for ease of presentation, we do not indicate the dependence of the amplitudes and phases on the spatial slow and fast scales)
%%%%%%%%%%%%%%%%%%%%%%%%%%%%%%%%%
\begin{eqnarray}
\label{polarization-L-2}
     P^{(1)}_{L,z}  (E_z^{(1)})
   & = &
   %%%%%%%%%%%%%%%%%%%%%%%%%%%%%%%%%%%%%%%%%%%%%%%%%%%%%
    \sum_{j=1}^{3}  \int_{-\infty}^{+\infty} \chi_{zz}^{(L)}(t-t_1) \left(A_{j}(\epsilon t_1) e^{i\theta_{j}( t_1 ) }
    +  B_{j}(\epsilon t_1) e^{-i\theta_{j}( t_1 ) } \right) d t_1
    \nonumber \\
    & = &
    %%%%%%%%%%%%%%%%%%%%%%%%%%%%%%%%%%%%%%%%%%%%%%%%%%%%%
%    \sum_{j=1}^{3}  \int_{-\infty}^{+\infty} \chi_{zz}^{(L)}(\tau) \left(A_{j}(\epsilon t - \epsilon\tau) e^{i\theta_{j}( t-\tau ) }
%    +  B_{j}(\epsilon t - \epsilon\tau) e^{-i\theta_{j}( t-\tau ) } \right) d \tau
%     \nonumber \\
%    & = &
    %%%%%%%%%%%%%%%%%%%%%%%%%%%%%%%%%%%%%%%%%%%%%%%%%%%%%
    \sum_{j=1}^{3}  \int_{-\infty}^{+\infty} \left( \chi_{zz}^{(L)}(\tau) e^{i\omega_j \tau} A_{j}(\epsilon t - \epsilon\tau) e^{i\theta_{j}( t) }
    + \chi_{zz}^{(L)}(\tau) e^{-i\omega_j \tau}  B_{j}(\epsilon t - \epsilon\tau) e^{-i\theta_{j}( t) } \right) d \tau \;.
    \nonumber \\ 
\end{eqnarray}
%%%%%%%%%%%%%%%%%%%%%%%%%%%%%%%%%
Next, we Taylor expand the amplitudes around the point $\epsilon t.$ We find
%%%%%%%%%%%%%%%%%%%%%%%%%%%%%%%%%
\begin{eqnarray}
\label{polarization-L-Taylor}
     P^{(1)}_{L,z}  (E_z^{(1)})
   & = &
    %%%%%%%%%%%%%%%%%%%%%%%%%%%%%%%%%%%%%%%%%%%%%%%%%%%%%
    \sum_{j=1}^{3}  \int_{-\infty}^{+\infty}  \chi_{zz}^{(L)}(\tau) e^{i\omega_j \tau} \left( A_{j}(T) -\epsilon \tau \partial_T A_j \right)e^{i\theta_{j}( t) } d\tau
      \nonumber \\
    & + &
    \sum_{j=1}^{3}  \int_{-\infty}^{+\infty}  \chi_{zz}^{(L)}(\tau) e^{-i\omega_j \tau} \left( B_{j}(T) -\epsilon \tau \partial_T B_j \right)e^{-i\theta_{j}( t) }d\tau
    +O(\epsilon^2) \;.
\end{eqnarray}
%%%%%%%%%%%%%%%%%%%%%%%%%%%%%%%%%
Define the Fourier transform
\begin{eqnarray}
\label{FT}
 \hat{\chi}_{zz}^{(L)}(\omega_j)  =  \int_{-\infty}^{+\infty}  \chi_{zz}^{(L)}(\tau) e^{i\omega_j \tau}  d\tau \;.
 \end{eqnarray}
%%%%%%%%%%%%%%%%%%%%%%%%%%%%%%%%
Using the Fourier representation of the susceptibility, Eq.~(\ref{polarization-L-Taylor}) now reads
%%%%%%%%%%%%%%%%%%%%%%%%%%%%%%%%%
\begin{equation}
\label{polarization-L-Taylor-1-a-b}
     P^{(1)}_{L,z}  (E_z^{(1)}) =  P^{(1,1)}_{L,z} (E_z^{(1)}) + \epsilon  P^{(1,2)}_{L,z} (E_z^{(1)}) +O(\epsilon^2) \;,
     \end{equation}
%%%%%%%%%%%%%%%%%%%%%%%%%%%%%%%%%
where
%%%%%%%%%%%%%%%%%%%%%%%%%%%%%%%
\begin{equation}
\label{polarization-L-Taylor-1-a}
     P^{(1,1)}_{L,z} (E_z^{(1)})
    = 
    \sum_{j=1}^{3}   \left( \hat{\chi}_{zz}^{(L)}(\omega_j) A_{j}(T) e^{i\theta_{j}( t) } 
    + \hat{\chi}_{zz}^{(L)}(-\omega_j) B_{j}(T) e^{-i\theta_{j}( t) } \right) \;,
  \end{equation}
  %%%%%%%%%%%%%%%%%%%%%%%%%%%%%%%
\begin{equation}
\label{polarization-L-Taylor-1-b}
   P^{(1,2)}_{L,z} (E_z^{(1)})
   = 
    \sum_{j=1}^{3}   \left( i \partial_\omega \hat{\chi}_{zz}^{(L)}(\omega_j) \partial_T A_{j}(T) e^{i\theta_{j}( t) } 
    + i \partial_\omega \hat{\chi}_{zz}^{(L)}(-\omega_j) \partial_T B_{j}(T) e^{-i\theta_{j}( t) } \right)
     \;.
\end{equation}
%%%%%%%%%%%%%%%%%%%%%%%%%%%%%%%%%%%%%%%%%%%%%%%%%%%%
Next, we turn our attention to the nonlinear polarization vector field to find its leading order contribution which is all that is needed for this calculation.
From the constraints on the nonlinear susceptibility tensor, the only non-zero component of the leading order 
vector $\mathbf{P}^{(NL)} (\mathbf{E}^{(1)})$ is given by
%%%%%%%%%%%%%%%%%%%%%%%%%%%%%%%%%
\begin{equation}
 \mathbf{P}^{(NL)} (\mathbf{E}^{(1)}) \left( \mathbf{r},t; \mathbf{R} , T\right) 
 = P^{(1)}_{NL,z} (E_z^{(1)}) \left( \mathbf{r}_{\perp},t; \mathbf{R}_{\perp} , T\right) \hat{\mathbf{z}} \;,
\end{equation}
%%%%%%%%%%%%%%%%%%%%%%%%%%%%%%%%%
where, 
%%%%%%%%%%%%%%%%%%%%%%%%%%%%%%%%%
\begin{equation}
    \label{polarization-NL-multiscale}
    P^{(1)}_{NL,z}  (E_z^{(1)}) \left( \mathbf{r}_{\perp},t; \mathbf{R}_{\perp} , T\right) 
   =  \int_{-\infty}^{+\infty} \chi_{zzz}^{(NL)}(t-t_1,t-t_2) 
   E^{(1)}_z  \left( \mathbf{r}_{\perp},t_1; \mathbf{R}_{\perp} , \epsilon t_1\right)  
   E^{(1)}_z  \left( \mathbf{r}_{\perp},t_2; \mathbf{R}_{\perp} , \epsilon t_2\right)  dt_1 dt_2 \;.
   %%%%%%%%%%%%%%%%%%%%%%%%%%%%%%%%%
\end{equation}
%%%%%%%%%%%%%%%%%%%%%%%%%%%%%%%%%%%%%%%%%%%%%%%%%%%%%%%%%%%%%%%%%
%%%%%%%%%%%%%%%%%%%%%%%%%%%%%%%%%
Substituting (\ref{six_wave_ansatz_E}) into (\ref{polarization-NL-multiscale}) we arrive at the following result (again, for notational purposes, we suppress the explicit dependence of the amplitudes and phases on the spatial slow and fast scales)
%%%%%%%%%%%%%%%%%%%%%%%%%%%%%%%%%
\begin{eqnarray}
\label{polarization-NL-2}
     P^{(1)}_{NL,z}  (E_z^{(1)})
   & = &
   %%%%%%%%%%%%%%%%%%%%%%%%%%%%%%%%%%%%%%%%%%%%%%%%%%%%%
    \sum_{n,m=1}^{3}  \int_{\mathbb{R}^2}  \chi_{zzz}^{(NL)}(t-t_1,t-t_2) 
     \left(A_{n}(\epsilon t_1) e^{i\theta_{n}( t_1 ) } +  B_{n}(\epsilon t_1) e^{-i\theta_{n}( t_1 ) } \right) 
     \nonumber \\
     && \left( A_{m}(\epsilon t_2) e^{i\theta_{m}( t_2 ) } +  B_{m}(\epsilon t_2) e^{-i\theta_{m}( t_2) } \right) 
     d t_1 dt_2
     %%%%%%%%%%%%%%%%%%%%%%%%%%%%%%%%%%%%%%%%%%%%%%%%%%%%%
      \;.
\end{eqnarray}
%%%%%%%%%%%%%%%%%%%%%%%%%%%%%%%%%
Let $\tau_1 = t-t_1$ and $\tau_2 = t-t_2.$ Note that $\theta_{n}( t_1) = \theta_{n}( t) + \omega_n \tau_1$ 
(similarly, $\theta_{m}( t_2) = \theta_{m}( t) + \omega_m \tau_2$). We now decompose the leading order nonlinear polarization into 
%%%%%%%%%%%%%%%%%%%%%%%%%%%%%%%%%
\begin{eqnarray}
\label{polarization-NL-decompose}
 P^{(1)}_{NL,z}  (E_z^{(1)}) =  P^{(1)}_{NL,res}  (E_z^{(1)}) + \text{non resonant terms} \;,
\end{eqnarray}
%%%%%%%%%%%%%%%%%%%%%%%%%%%%%%%%%
where $P^{(1)}_{NL,res}  (E_z^{(1)})$ is the remaining contribution from $ P^{(1)}_{NL,z}(E_z^{(1)})$ that contain resonant terms. 
From (\ref{polarization-NL-2}) we have
%%%%%%%%%%%%%%%%%%%%%%%%%%%%%%%%%
\begin{align}
\label{polarization-NL-3-res}
    P^{(1)}_{NL,res} (E_z^{(1)})
   & = 
   %%%%%%%%%%%%%%%%%%%%%%%%%%%%%%%%%%%%%%%%%%%%%%%%%%%%%
    \sum_{n,m=1}^{3}  \int_{\mathbb{R}^2}  \chi_{zzz}^{(NL)}(\tau_1,\tau_2) e^{i(\omega_n\tau_1 + \omega_m\tau_2)} 
     A_{n}(\epsilon t - \epsilon\tau_1) A_{m}(\epsilon t- \epsilon\tau_2) e^{i(\theta_{n}( t ) +\theta_m(t))} d \tau_1 d\tau_2
     \nonumber \\
       & + 
       %%%%%%%%%%%%%%%%%%%%%%%%%%%%%%%%%%%%%%%%%%%%%%%%%%%%%
    \sum_{n,m=1}^{3}  \int_{\mathbb{R}^2}  \chi_{zzz}^{(NL)}(\tau_1,\tau_2) e^{-i(\omega_n\tau_1 + \omega_m\tau_2)} 
     B_{n}(\epsilon t - \epsilon\tau_1) B_{m}(\epsilon t- \epsilon\tau_2) e^{-i(\theta_{n}( t ) +\theta_m(t))} d \tau_1 d\tau_2\;.
     %%%%%%%%%%%%%%%%%%%%%%%%%%%%%%%%%%%%%%%%%%%%%%%%%%%%%
       \nonumber \\
\end{align}
%%%%%%%%%%%%%%%%%%%%%%%%%%%%%%%%%
From (\ref{Maxwell_vector-1-asym-3}) and $\mathbf{P}^{(NL)} (\mathbf{E}^{(1)})$, 
it is sufficient to consider only the leading order contribution of (\ref{polarization-NL-3-res}).
%%%%%%%%%%%%%%%%%%%%%%%%%%%%%%%%%
%%%%%%%%%%%%%%%%%%%%%%%%%%%%%%%%%
\begin{eqnarray}
\label{polarization-NL-4-res}
    P^{(1)}_{NL,res} (E_z^{(1)})
   & = &
   %%%%%%%%%%%%%%%%%%%%%%%%%%%%%%%%%%%%%%%%%%%%%%%%%%%%%
    \sum_{n,m=1}^{3}   \hat{\chi}_{zzz}^{(NL)}(\omega_n,\omega_m) 
     A_{n}(T) A_{m}(T) e^{i(\theta_{n}( t ) +\theta_m(t))} 
     \nonumber \\
       & + &
       %%%%%%%%%%%%%%%%%%%%%%%%%%%%%%%%%%%%%%%%%%%%%%%%%%%%%
  \sum_{n,m=1}^{3}   \hat{\chi}_{zzz}^{(NL)}(-\omega_n,-\omega_m) 
     B_{n}(T) B_{m}(T) e^{-i(\theta_{n}( t ) +\theta_m(t))} +O(\epsilon) \;,
     %%%%%%%%%%%%%%%%%%%%%%%%%%%%%%%%%%%%%%%%%%%%%%%%%%%%%
\end{eqnarray}
%%%%%%%%%%%%%%%%%%%%%%%%%%%%%%%%%
where $ \hat{\chi}_{zzz}^{(NL)}$ denotes the two-dimensional Fourier transform of $\chi_{zzz}^{(NL)}$ defined by
%%%%%%%%%%%%%%%%%%%%%%%%%%%%%%%%%
\begin{eqnarray}
\label{2D-FT}
 \hat{\chi}_{zzz}^{(NL)} (\omega_n,\omega_m)  
 =   \int_{\mathbb{R}^2}  \chi_{zzz}^{(NL)}(\tau_1,\tau_2) e^{i(\omega_n\tau_1 + \omega_m\tau_2)} d \tau_1 d\tau_2\ \;.
 \end{eqnarray}
%%%%%%%%%%%%%%%%%%%%%%%%%%%%%%%%
Having identified the leading and order $\epsilon$ contributions arising from the linear and nonlinear polarization vectors, 
we next turn our focus on separating the space-time scales in (\ref{Maxwell_vector-1-asym-3}) and (\ref{Maxwell_vector-2-asym-3}). 
The leading order equations are given by:
%%%%%%%%%%%%%%%%%%%%%%%%%%%%%%%%%%%%%%%%%%%%%%%%%%%%%%%%%%%%%%%%%%
\begin{align}
    \label{Maxwell_vector-1-asym-leading}
   &  \frac{1}{c^2} \frac{ \partial^2}{\partial t^2}  \left(   \mathbf{E}^{(1)} + P^{(1,1)}_{L,z} (E_z^{(1)}) \hat{\mathbf{z}} \right)
   +  \nabla_\mathbf{r} ( \nabla_\mathbf{r} \cdot\mathbf{E}^{(1)}) -   
   \nabla_\mathbf{r}^{2}\mathbf{E}^{(1)}    = 0 \;,
 \\ \nonumber
   \\
  \label{Maxwell_vector-2-asym-leading}
  & \nabla_\mathbf{r} \cdot \left(   \mathbf{E}^{(1)} + P^{(1,1)}_{L,z} (E_z^{(1)}) \hat{\mathbf{z}}   \right) 
  = 0\;.
\end{align}
%%%%%%%%%%%%%%%%%%%%%%%%%%%%%%%%%%%%%%%%%%%%%%%%%%%%%%%%%%%%%%%%%%
In the $\hat{\mathbf{x}}$ and $\hat{\mathbf{y}}$ directions, (\ref{Maxwell_vector-1-asym-leading}) is automatically satisfied. 
Furthermore, Eq.~(\ref{Maxwell_vector-2-asym-leading}) leads to no new information since both $\mathbf{E}^{(1)}$ 
and $P^{(1,1)}_{L,z}  \hat{\mathbf{z}}$ are dependent only on $\mathbf{r}_\perp$ and has a non-zero contribution only in the $\hat{\mathbf{z}}$ component.
With this at hand, the nontrivial part of Eq.~(\ref{Maxwell_vector-1-asym-leading}) read
%%%%%%%%%%%%%%%%%%%%%%%%%%%%%%%%%
\begin{align}
    \label{Maxwell_vector-1a}
    %%%%%%%%%%%%%%%%%%%%%%%%%%%%%%
   &   \mathcal{L} \left(  E^{(1)}_z , P_{L,z}^{(1,1)}(E_z^{(1)}) \right) \equiv
    \frac{1}{c^2} \frac{ \partial^2}{\partial t^2}  \left( E^{(1)}_z + P_{L,z}^{(1,1)} (E_z^{(1)})  \right) 
   -  \nabla_\mathbf{r_{\perp}}^{2} E^{(1)}_z= 0 \;.
\end{align}
%%%%%%%%%%%%%%%%%%%%%%%%%%%%%%%%%
%%%%%%%%%%%%%%%%%%%%%%%%%%%%%%%%%
Substituting (\ref{six_wave_ansatz_E}) and (\ref{polarization-L-Taylor-1-a})  into \eqref{Maxwell_vector-1a} leads to the dispersion relation (in an implicit form)
%%%%%%%%%%%%%%%%%%%%%%%%%%%%%%%%%
\begin{equation}
    \label{disp_E}
     | \mathbf{k} |^{2}  = \frac{\omega^2}{c^2} \left( 1 +  \hat{\chi}_{zz}^{(L)}(\omega) \right)\,.
\end{equation}
%%%%%%%%%%%%%%%%%%%%%%%%%%%%%%%%%
The group velocity associated with the above implicit dispersion relation is given by
%%%%%%%%%%%%%%%%%%%%%%%%%%%%%%%%%
\begin{equation}
    \label{group-velocity}
 \mathbf{V}_g  \equiv  \nabla_ \mathbf{k} \omega 
 = \frac{c^2 \mathbf{k} }{\omega \left( 1 + \hat{\chi}_{zz}^{(L)} + \frac{\omega}{2}\partial_\omega \hat{\chi}_{zz}^{(L)}  \right)}  \,.
\end{equation}
%%%%%%%%%%%%%%%%%%%%%%%%%%%%%%%%%
For example, in \cite{AS_Book} (and the references therein), the dispersion relation   
%%%%%%%%%%%%%%%%%%%%%%%%%%%%%%%%%
\begin{equation}
    \label{disp_E1}
      \hat{\chi}_{zz}^{(L)}(\omega) =\frac{\Omega^2}{\omega_0^2 - \omega^2}\,,
\end{equation}
%%%%%%%%%%%%%%%%%%%%%%%%%%%%%%%%%
where $\omega_0, \Omega$ are constants was shown to have triad resonance satisfying Eq.~(\ref{resonance}).
We turn our focus on the order $\epsilon$ contributions that arise from Eq.~(\ref{Maxwell_vector-1-asym-3}). Since $\mathbf{E}^{(1)}, \mathbf{P}^{(L)} (\mathbf{E}^{(1)})$ and $\mathbf{P}^{(NL)} (\mathbf{E}^{(1)})$ are independent of $z, Z$ and all point in the $\hat{\mathbf{z}}$ direction, then to order $\epsilon$, one can self-consistently assume that the vector fields
$\mathbf{E}^{(2)}, \mathbf{P}^{(L)} (\mathbf{E}^{(2)})$ also $z, Z$ independent and are parallel to the 
$z$ axis. As such, we arrive at the order $\epsilon$ equation
%%%%%%%%%%%%%%%%%%%%%%%%%%%%%%%%%
\begin{align}
    \label{polarization_4}
    -   \mathcal{L} \left(  E^{(2)}_z , P_{L,z}^{(1,1)}(E_z^{(2)}) \right) 
    & = 
  \frac{1}{c^2}\left[ 2\partial_{tT}\left( E^{(1)}_z + P_{L,z}^{(1,1)}(E_z^{(1)}) \right) + \partial^2_{t}  P^{(1,2)}_{L,z} (E_z^{(1)})
  + \partial^2_{t}  P^{(1)}_{NL,res} (E_z^{(1)}) \right] 
   \nonumber \\
  &-
     2 \left(\nabla_{\mathbf{R}_{\perp}} \cdot \nabla_{\mathbf{r}_{\perp}}\right) E_z^{(1)} \;.
\end{align}
%%%%%%%%%%%%%%%%%%%%%%%%%%%%%%%%%
In addition, the order $\epsilon$ contribution that comes from (\ref{Maxwell_vector-2-asym-3}) is given by
%%%%%%%%%%%%%%%%%%%%%%%%%%%%%%%%%
\begin{equation}
 \label{Maxwell_vector-2-asym-3-epsilon}
-  \nabla_\mathbf{r} \cdot \left(   \mathbf{E}^{(2)} + \mathbf{P}^{(L,1)} (\mathbf{E}^{(2)})   \right) 
  =
    \nabla_{\mathbf{R}} \cdot \left(   \mathbf{E}^{(1)} +   \mathbf{P}^{(L,1)} ( \mathbf{E}^{(1)}) \right) 
  + \nabla_\mathbf{r} \cdot  \mathbf{P}^{(NL,1)} (\mathbf{E}^{(1)}) 
  \;.
\end{equation}
%%%%%%%%%%%%%%%%%%%%%%%%%%%%%%%%%
In (\ref{Maxwell_vector-2-asym-3-epsilon}), $\mathbf{P}^{(L,1)}$ and $\mathbf{P}^{(NL,1)}$ are the leading order terms in the $\epsilon$ 
expansion of $\mathbf{P}^{(L)}$ and $\mathbf{P}^{(NL)}$ respectively. With the ansatz (\ref{six_wave_ansatz_E}), it is evident that the right hand side of 
(\ref{Maxwell_vector-2-asym-3-epsilon}) vanishes (since it is independent of both the fast and slow $(z,Z)$ and points in the $\hat{\mathbf{z}}$ direction). 
Thus, under the above assumption that both $\mathbf{E}^{(2)}$ and $\mathbf{P}^{(L)} (\mathbf{E}^{(2)})$ be also $z, Z$ independent (and parallel to the $z$ axis), Eq.~(\ref{Maxwell_vector-2-asym-3-epsilon}) leads to no new constraints on the electric and polarization fields. 
Next, substituting Eqns.~(\ref{six_wave_ansatz_E}), (\ref{polarization-L-Taylor-1-a}), (\ref{polarization-L-Taylor-1-b}) and (\ref{polarization-NL-4-res}) into (\ref{polarization_4}) leads to (for simplicity, we suppress the explicit dependence of the phases $\theta_j$ and amplitudes $A_j , B_j$ on the 
spatio-temporal fast and slow variables)
%%%%%%%%%%%%%%%%%%%%%%%%%%%%%%%%%%%%%%%%%%%%%%%%%%%%%%%%%%%%%%%
%%%%%%%%%%%%%%%%%%%%%%%%%%%%%%%%%%%%%%%%%%%%%%%%%%%%%%%%%%%%%%%
\begin{align}
    \label{six-waves-simple}
    %%%%%%%%%%%%%%%%%%%%%%%%
     c^2 \mathcal{L} \left(  E^{(2)}_z , P_{L,z}^{(1,1)}(E_z^{(2)}) \right) 
&=
  \sum_{j=1}^{3} \left[ 2i\omega_j  \left( 1 +  \hat{\chi}_{zz}^{(L)}(\omega_j) 
  +  \frac{\omega_j}{2} \partial_\omega \hat{\chi}_{zz}^{(L)}(\omega_j)  \right) \frac{\partial A_{j}}{\partial T} 
     +  2i c^2(\mathbf{k}_j  \cdot \nabla_{\mathbf{R}_{\perp}} ) A_j \right] e^{i\theta_j}
        %%%%%%%%%%%%%%%%%%%%%
         \nonumber \\
    &-
   \sum_{j=1}^{3} \left[ 2i\omega_j \left( 1 +  \hat{\chi}_{zz}^{(L)}(-\omega_j) 
  +  \frac{\omega_j}{2} \partial_\omega \hat{\chi}_{zz}^{(L)}(-\omega_j)  \right) \frac{\partial B_{j}}{\partial T} 
     +  2i c^2 (\mathbf{k}_j  \cdot \nabla_{\mathbf{R}_{\perp}} ) B_j \right] e^{-i\theta_j}
        %%%%%%%%%%%%%%%%%%%%%
     \nonumber \\
    &+
    %%%%%%%%%%%%%%%%%%%%%
      %%%%%%%%%%%%%%%%%%%%%%%%%%%%%%%%%%%%%%%%%%%%%%%%%%%%%
    \sum_{n,m=1}^{3} (\omega_n + \omega_m )^2  \hat{\chi}_{zzz}^{(NL)}(\omega_n,\omega_m) 
     A_{n} A_{m} e^{i(\theta_{n} +\theta_m)} 
     \nonumber \\
       & +
       %%%%%%%%%%%%%%%%%%%%%%%%%%%%%%%%%%%%%%%%%%%%%%%%%%%%%
  \sum_{n,m=1}^{3} (\omega_n + \omega_m )^2 \hat{\chi}_{zzz}^{(NL)}(-\omega_n,-\omega_m) 
     B_{n} B_{m} e^{-i(\theta_{n} +\theta_m)} \;.
     %%%%%%%%%%%%%%%%%%%%%%%%%%%%%%%%%%%%%%%%%%%%%%%%%%%%
\end{align}
%%%%%%%%%%%%%%%%%%%%%%%%%%%%%%%%
%%%%%%%%%%%%%%%%%%%%%%%%%%%%%%%%%
%%%%%%%%%%%%%%%%%%%%%%%%%%%%%%%%%%%%%%%%%%%%%%%%%%%%%%%%%%%%%%%%%%%%
Finally, collecting resonant terms that lead to unbounded growth (recall that resonance occurs whenever the 
condition $\theta_1 + \theta_2 + \theta_3 =0$ is satisfied), we arrive at the following six-wave resonance interaction equations (see Appendix for more details):
%%%%%%%%%%%%%%%%%%%%%%%%%%%%%%%%%%%%%%%%%%%%%%%%%%%%%%%%%%%%%%%%%%%%%
\bse
   \label{6wWaveEqA}
%%%%%%%%%%%%%%%%%%%%%%%%%%%%%%%%%
\begin{eqnarray}
%%%%%%%%%%%%%%%%%%%%%%%%%%%%%%%%%%%%%%%%%%%%%%%%%%%%%%%%
    \partial_{T} A_1 + (\mathbf{V}_g^{(1)} \cdot \nabla_{\mathbf{R}_{\perp}}) A_1 &=& i\gamma_{1}B_{2} B_{3} \,,\\
    %%%%%%%%%%%%%%%%%%%%%%%%%%%%%%%%%%%%%%%%%%%%%%%%%%%%%%%
    \partial_{T} A_2 + (\mathbf{V}_g^{(2)} \cdot \nabla_{\mathbf{R}_{\perp}}) A_2 &=& i\gamma_{2}B_{3} B_{1} \,,\\
    %%%%%%%%%%%%%%%%%%%%%%%%%%%%%%%%%%%%%%%%%%%%%%%%%%%%%%
   \partial_{T} A_3 + (\mathbf{V}_g^{(3)} \cdot \nabla_{\mathbf{R}_{\perp}}) A_3 &=& i\gamma_{3}B_{1} B_{2} \,,\\
    %%%%%%%%%%%%%%%%%%%%%%%%%%%%%%%%%%%%%%%%%%%%%%%%%%%%%%
    %%%%%%%%%%%%%%%%%%%%%%%%%%%%%%%%%%%%%%%%%%%%%%%%%%%%%%%%
    \partial_{T} B_1 + (\mathbf{V}_g^{(1)} \cdot \nabla_{\mathbf{R}_{\perp}}) B_1 &=& - i\gamma_{1}A_{2} A_{3} \,,\\
    %%%%%%%%%%%%%%%%%%%%%%%%%%%%%%%%%%%%%%%%%%%%%%%%%%%%%%%
    \partial_{T} B_2 + (\mathbf{V}_g^{(2)} \cdot \nabla_{\mathbf{R}_{\perp}}) B_2 &=& -i\gamma_{2}A_{3} A_{1} \,,\\
    %%%%%%%%%%%%%%%%%%%%%%%%%%%%%%%%%%%%%%%%%%%%%%%%%%%%%%
   \partial_{T} B_3 + (\mathbf{V}_g^{(3)} \cdot \nabla_{\mathbf{R}_{\perp}}) B_3 &=& -i\gamma_{3}A_{1} A_{2} \,,
    %%%%%%%%%%%%%%%%%%%%%%%%%%%%%%%%%%%%%%%%%%%%%%%%%%%%%%
    %%%%%%%%%%%%%%%%%%%%%%%%%%%%%%%%%%%%%%%%%%%%%%%%%%%%%%
   \label{6wWaveEqB}
\end{eqnarray}
\ese
%%%%%%%%%%%%%%%%%%%%%%%%%%%%%%%%%
where, 
%%%%%%%%%%%%%%%%%%%%%%%%%%%%%%%%%
%%%%%%%%%%%%%%%%%%%%%%%%%%%%%%%%%
\begin{equation}
    \label{group-velocity-j}
 \mathbf{V}^{(j)}_g  \equiv  
  \frac{c^2}{\omega_j} \left( 1 + \hat{\chi}_{zz}^{(L)} (\omega_j) 
  + \frac{\omega_j}{2}\partial_{\omega_j} \hat{\chi}_{zz}^{(L)} (\omega_j)  \right)^{-1}  \mathbf{k}_j \,,\;\;\; j=1,2,3\;,
\end{equation}
%%%%%%%%%%%%%%%%%%%%%%%%%%%%%%%%%
and
%%%%%%%%%%%%%%%%%%%%%%%%%%%%%%%%%
\begin{equation}
    \label{gamma-j}
    \gamma_1 = \omega_1 \hat{\chi}_{zzz}^{(NL)}(\omega_2,\omega_3)\;,\;\;\;\;\;
    \gamma_2 = \omega_2 \hat{\chi}_{zzz}^{(NL)}(\omega_1,\omega_3)\;,\;\;\;\;\;
    \gamma_3 = \omega_3 \hat{\chi}_{zzz}^{(NL)}(\omega_1,\omega_2)\;
    \end{equation}
  where we have assumed: 
$\hat{\chi}_{zzz}^{(NL)}(-\omega_1,-\omega_2) = \hat{\chi}_{zzz}^{(NL)}(\omega_1,\omega_2)$
  and $\hat{\chi}_{zzz}^{(NL)}(\omega_1,\omega_2) = \hat{\chi}_{zzz}^{(NL)}(\omega_2,\omega_1)$.
  
%%%%%%%%%%%%%%%%%%%%%%%%%%%%%%%%%%%%%%%%%%%%%%%%%%%%%%%%%%%%%%%%%%
\subsection{Laser model}
%%%%%%%%%%%%%%%%%%%%%%%%%%%%%%%%%%%%%%%%%%%%%%%%%%%%%%%%%%%%%%%%%%
In this section, we provide an alternative derivation of the six-wave interaction equations starting from a model commonly used in laser optics  (see \cite{AS_Book} and related references therein). The starting point is the coupled system (\ref{Maxwell_vector-1}) and (\ref{Maxwell_vector-2}) that govern
the dynamics of the electric field. Unlike the case discussed in Sec.~(3.1) where the polarization vector is related to the electric field via a convolution integral; here the polarization field is determined from a dynamical model used in lasers. Specifically, we will assume that the polarization component 
$P_j$ obeys the equation
%%%%%%%%%%%%%%%%%%%%%%%%%%%%%%%%%
\begin{equation}
    \label{polarization-laser}
    \partial_{t}^{2}P_{j}+\omega_{0}^{2}P_{j}+\sum_{k,\ell =1}^{3}d_{jk\ell}P_{k}P_{\ell} - \frac{\Omega^{2}}{\mu_{0}c^{2}}E_{j} =0\,,\;\;\; j=1,2,3\,,
\end{equation}
%%%%%%%%%%%%%%%%%%%%%%%%%%%%%%%%%
where $d_{jk\ell}=d_{j\ell k}$, $\omega_{0}$, and $\Omega$ are inherent to the given medium. We shall follow similar assumptions made in Sec.~(3.1) regarding the electric and polarization fields. That is to say, we let both $\mathbf{E}$ and $\mathbf{P}$ be $z$ independent and have their third component be the only nonzero element. With the notation $\mathbf{E} = E \hat{\mathbf{z}}$ and  $\mathbf{P} = P \hat{\mathbf{z}}$ we arrive at the quadratically nonlinear coupled  system:
%%%%%%%%%%%%%%%%%%%%%%%%%%%%%%%%%
\begin{equation}
    \label{Maxwell_E-linear-laser}
    \partial_{t}^{2} E - c^{2}\nabla^{2} E + \mu_{0}c^{2}\partial_{t}^{2}P  =0\,,
\end{equation}
%%%%%%%%%%%%%%%%%%%%%%%%%%%%%%%%%
%%%%%%%%%%%%%%%%%%%%%%%%%%%%%%%%%
\begin{equation}
    \label{Maxwell_P-linear-laser}
    \partial_{t}^{2} P  +\omega_0^2 P + d_{zzz}P^2 - \frac{\Omega^2}{\mu_0 c^2}E =0 \,,
\end{equation}
%%%%%%%%%%%%%%%%%%%%%%%%%%%%%%%%%
where the notation (from optics) $d_{333}\equiv d_{zzz}$ has been adopted. Next, introduce the slow time and space variables (\ref{perturb}) and the perturbation expansions
%%%%%%%%%%%%%%%%%%%%%%%%%%%%%%%%%
\begin{equation}
\label{perturb-laser}
     E=\epsilon E_{1}+\epsilon^{2}E_{2}+\cdots\,,\;\;\; 
     P=\epsilon P_{1}+\epsilon^2P_{2}+\cdots\,.
\end{equation}
%%%%%%%%%%%%%%%%%%%%%%%%%%%%%%%%%
%%%%%%%%%%%%%%%%%%%%%%%%%%%%%%%%%
Substituting (\ref{perturb-laser}) into (\ref{Maxwell_E-linear-laser}) and (\ref{Maxwell_P-linear-laser}), we arrive at the leading order equation
%%%%%%%%%%%%%%%%%%%%%%%%%%%%%%%%%%%%%%%%%%%%%%%%
%%%%%%%%%%%%%%%%%%%%%%%%%%%%%%%%%
\begin{equation}
\mathbb{L} \begin{bmatrix}
         E_1\\\\
       P_1 
    \end{bmatrix}    
    =0\;,\;\;\;\;\;\;\;
%%%%%%%%%%%%%%%%%%%%%%%%%%%%%%%%
    \mathbb{L} \equiv \begin{bmatrix}
         \partial^2_t - c^{2} \nabla^2_ \mathbf{r} &&\mu_{0}c^{2} \partial^2_t\\ \\
        - \frac{\Omega^{2}}{\mu_{0}c^{2}} && \partial^2_t +\omega_0^2
    \end{bmatrix}    \;.
\end{equation}
%%%%%%%%%%%%%%%%%%%%%%%%%%%%%%%%%
To determine the dispersion relation, we seek a plane wave solution in the form
%%%%%%%%%%%%%%%%%%%%%%%%%%%%%%%%%
\begin{equation}
    \label{plane_wave-0-laser}
    E_{1} = \mathcal{E}_1 e^{i\theta}\,,\;\;\; P_1 = \mathcal{P}_{1}  e^{i\theta}\,,\;\;\;\theta=\mathbf{k}\cdot \mathbf{r}-\omega t\,.
\end{equation}
%%%%%%%%%%%%%%%%%%%%%%%%%%%%%%%%%
Then we find
%%%%%%%%%%%%%%%%%%%%%%%%%%%%%%%%%
\begin{equation}
    \label{disp-1-laser}
    c^{2} | \mathbf{k} |^2 = \omega^{2}-\frac{\omega^{2}\Omega^{2}}{\omega^{2}-\omega_{0}^{2}}\,,\;\;\;\;\;
    \mathbf{V}_g =   \nabla_{\mathbf{k}}\omega 
   =
   \frac{c^2 \mathbf{k}}{\omega \left( 1 + \frac{\omega_0^{2}\Omega^{2}}{ (\omega_0^{2}-\omega^2 )^2} \right)} \,.
\end{equation}
%%%%%%%%%%%%%%%%%%%%%%%%%%%%%%%%%
%%%%%%%%%%%%%%%%%%%%%%%%%%%%%%%%%
The corresponding eigenvector is given by:
%%%%%%%%%%%%%%%%%%%%%%%%%%%%%%%%
\begin{equation} 
%%%%%%%%%%%%
     \begin{bmatrix}
         \mathcal{E}_1\\\\
       \mathcal{P}_1
    \end{bmatrix}    
    = 
    \begin{bmatrix}
         1\\\\
       \frac{\Omega^{2}}{\mu_0 c^2 (\omega_0^{2}-\omega^{2})} 
    \end{bmatrix}    
    \;.
\end{equation}
%%%%%%%%%%%%%%%%%%%%%%%%%%%%%%%%%%
The following adjoint problem will be later used to determine the six-wave interaction equations:
%%%%%%%%%%%%%%%%%%%%%%%%%%%%%%%%
\begin{equation}
\underbrace{
    \begin{bmatrix}
         -\omega^2 + c^{2} | \mathbf{k} |^2&  - \frac{\Omega^{2}}{\mu_{0}c^{2}}\\ \\
 -\mu_{0}c^{2} \omega^2       & -\omega^2 +\omega_0^2
    \end{bmatrix}}_{\mathbb{L}^{A}}    
\mathbf{w}^A=0\;,
\;\;\;\;\;
 \mathbf{w}^A   
    = 
    \begin{bmatrix}
         1\\\\
       \frac{\mu_0 c^2\omega^2}{(\omega_0^{2}-\omega^{2})}
    \end{bmatrix}    \;,
  \end{equation}
%%%%%%%%%%%%%%%%%%%%%%%%%%%%%%%%%%
%%%%%%%%%%%%%%%%%%%%%%%%%%%%%%%%%
where $\mathbf{w}^A$ denotes the adjoint eigenvector.
%%%%%%%%%%%%%%%%%%%%%%%%%%%%%%%%
%%%%%%%%%%%%%%%%%%%%%%%%%%%%%%%%
The equation at order $\epsilon$ reads
%%%%%%%%%%%%%%%%%%%%%%%%%%%%%%%%%%%%%%%%%%%%%%%%
%%%%%%%%%%%%%%%%%%%%%%%%%%%%%%%%%
\begin{equation}
\label{order-epsilon-eqn}
    \mathbb{L} \begin{bmatrix}
         E_2\\\\
       P_2 
    \end{bmatrix}    
    = 
    \mathcal{F}\;,\;\;\;\;\;\;\;
    \mathcal{F}
    =
    \begin{bmatrix}
         -2 \partial_{tT} E_{1}
   - 2\mu_{0}c^{2} \partial_{tT} P_{1}
    + 2c^{2}  \left( \nabla_ \mathbf{r} \cdot \nabla_{\mathbf{R}} \right) E_1
      \\\\
      -2\partial_{tT}  P_{1} -P_1^2
    \end{bmatrix}    \;.
\end{equation}
%%%%%%%%%%%%%%%%%%%%%%%%%%%%%%%%
We now make a six-wave ansatz in the form given by (\ref{six_wave_ansatz_E}) for the electric field and
%%%%%%%%%%%%%%%%%%%%%%%%%%%%%%%%%
%%%%%%%%%%%%%%%%%%%%%%%%%%%%%%%%%
\begin{equation}
\label{six_wave_ansatz_P}
    P_1 \left( \mathbf{r}_{\perp},t; \mathbf{R}_{\perp} , T\right) 
    =
     \frac{\Omega^{2}}{\mu_0 c^2} 
    \sum_{j=1}^{3}  \frac{1}{ (\omega_0^{2}-\omega_j^2)}  \Big[ A_{j}(\mathbf{R}_{\perp} , T) e^{i\theta_{j}( \mathbf{r}_{\perp},t ) }
    + B_{j}(\mathbf{R}_{\perp} , T) e^{-i\theta_{j}( \mathbf{r}_{\perp},t ) }\Big]\,.
\end{equation}
%%%%%%%%%%%%%%%%%%%%%%%%%%%%%%%%%
Taking the square leads to
%%%%%%%%%%%%%%%%%%%%%%%%%%%%%%%%%
%%%%%%%%%%%%%%%%%%%%%%%%%%%%%%%%%
\begin{align}
\label{six_wave_ansatz_P-2}
    P^2_1 
    =
     \frac{2\Omega^{4}}{\mu^2_0 c^4} 
    \sum_{j=1}^{3}\left(  \mathbb{P}^{(1)}_j e^{i\theta_j } +  \mathbb{P}^{(2)}_j e^{-i\theta_j }\right)    + \text{non resonant terms} \,,
\end{align}
%%%%%%%%%%%%%%%%%%%%%%%%%%%%%%%%%
where $\mathbb{P}^{(k)}_j,  \;j=1,2,3, \;k=1,2$ depends on the frequencies and the amplitudes $A_j, B_j$ given by 
%%%%%%%%%%%%%%%%%%%%%%%%%%%%%%%%%
\begin{align}
\label{P-1-2-coeff-B}
%%%%%%%%%%%%%%%%%%%%%%%%%%%%%%%%%%%%%%%%%%%%%%%%%%%%%%%%%%%%%%%
    \mathbb{P}^{(1)}_1 =  \frac{B_2 B_3}{(\omega_{0}^{2} - \omega_{2}^2)(\omega_{0}^{2} - \omega_{3}^2)} \;,\;\;\;\;\;
    %%%%%%%%%%%%%%%%%%%%%%%%%%%%%%%%%%%%%%%%%%%%%%%%%%%%%%%%%%%%%%%
      \mathbb{P}^{(1)}_2 =  \frac{B_1 B_3}{(\omega_{0}^{2} - \omega_{1}^2)(\omega_{0}^{2} - \omega_{3}^2)} \;,\;\;\;\;\;
      %%%%%%%%%%%%%%%%%%%%%%%%%%%%%%%%%%%%%%%%%%%%%%%%%%%%%%%%%%%%%%%
        \mathbb{P}^{(1)}_3 =  \frac{B_1 B_2}{(\omega_{0}^{2} - \omega_{1}^2)(\omega_{0}^{2} - \omega_{2}^2)} \;,
  %%%%%%%%%%%%%%%%%%%%%%%%%%%%%%%%%%%%%%%%%%%%%%%%%%%%%%%%%%%%%%%
\end{align}
%%%%%%%%%%%%%%%%%%%%%%%%%%%%%%%%%
%%%%%%%%%%%%%%%%%%%%%%%%%%%%%%%%%
%%%%%%%%%%%%%%%%%%%%%%%%%%%%%%%%%
\begin{align}
\label{P-1-2-coeff-A}
%%%%%%%%%%%%%%%%%%%%%%%%%%%%%%%%%%%%%%%%%%%%%%%%%%%%%%%%%%%%%%%
    \mathbb{P}^{(2)}_1 =  \frac{A_2 A_3}{(\omega_{0}^{2} - \omega_{2}^2)(\omega_{0}^{2} - \omega_{3}^2)} \;,\;\;\;\;\;
    %%%%%%%%%%%%%%%%%%%%%%%%%%%%%%%%%%%%%%%%%%%%%%%%%%%%%%%%%%%%%%%
      \mathbb{P}^{(2)}_2 =  \frac{A_1 A_3}{(\omega_{0}^{2} - \omega_{1}^2)(\omega_{0}^{2} - \omega_{3}^2)} \;,\;\;\;\;\;
      %%%%%%%%%%%%%%%%%%%%%%%%%%%%%%%%%%%%%%%%%%%%%%%%%%%%%%%%%%%%%%%
        \mathbb{P}^{(2)}_3 =  \frac{A_1 A_2}{(\omega_{0}^{2} - \omega_{1}^2)(\omega_{0}^{2} - \omega_{2}^2)} \;.
  %%%%%%%%%%%%%%%%%%%%%%%%%%%%%%%%%%%%%%%%%%%%%%%%%%%%%%%%%%%%%%%
\end{align}
%%%%%%%%%%%%%%%%%%%%%%%%%%%%%%%%%
We now decompose the vector $\mathcal{F}$ as 
%%%%%%%%%%%%%%%%%%%%%%%%%%%%%%%%%
\begin{equation}
\label{vector-F}
    \mathcal{F}
    =
     \sum_{j=1}^{3}\left(   \mathcal{F}^{(1)}_j e^{i\theta_j } + \mathcal{F}^{(2)}_j e^{-i\theta_j }\right) 
      \;,
\end{equation}
%%%%%%%%%%%%%%%%%%%%%%%%%%%%%%%%
where
%%%%%%%%%%%%%%%%%%%%%%%%%%%%%%
%%%%%%%%%%%%%%%%%%%%%%%%%%%%%%%%%
\begin{equation}
\label{vector-F-1}
  \mathcal{F}^{(1)}_j 
    =
    \begin{bmatrix}
         2i\omega_j \partial_{T} A_j
  + \frac{2i\omega_j   \Omega^{2}}{ (\omega_0^{2}-\omega_j^{2})}  \partial_{T} A_j
    + 2ic^{2}  \left( \mathbf{k}_j \cdot \nabla_{\mathbf{R}} \right) A_j
      \\\\
       \frac{2i\omega_j \Omega^{2}}{\mu_0 c^2 (\omega_0^{2}-\omega_j^{2})}  \partial_{T}  A_j
       -  \frac{2\Omega^{4}}{\mu^2_0 c^4}  \mathbb{P}^{(1)}_j 
    \end{bmatrix}    
  \;,
\end{equation}
%%%%%%%%%%%%%%%%%%%%%%%%%%%%%%%%
and
%%%%%%%%%%%%%%%%%%%%%%%%%%%%%%%%%
\begin{equation}
\label{vector-F-2}
  \mathcal{F}^{(2)}_j 
    =
    \begin{bmatrix}
       -  2i\omega_j \partial_{T} B_j
  - \frac{2i\omega_j   \Omega^{2}}{ (\omega_0^{2}-\omega_j^{2})}  \partial_{T} B_j
    - 2ic^{2}  \left( \mathbf{k}_j \cdot \nabla_{\mathbf{R}} \right) B_j
      \\\\
      - \frac{2i\omega_j \Omega^{2}}{\mu_0 c^2 (\omega_0^{2}-\omega_j^{2})}  \partial_{T}  B_j
       -  \frac{2\Omega^{4}}{\mu^2_0 c^4}  \mathbb{P}^{(2)}_j 
    \end{bmatrix}    
  \;.
\end{equation}
%%%%%%%%%%%%%%%%%%%%%%%%%%%%%%%%
From the Fredholm alternative, the solvability condition associated with (\ref{order-epsilon-eqn}) is expressed as
%%%%%%%%%%%%%%%%%%%%%%%%%%%%%%%%%
%%%%%%%%%%%%%%%%%%%%%%%%%%%%%%%%%
\begin{equation}
\label{order-epsilon-eqn-Frdholm}
  \langle
   \mathbf{w}^A  ,  \mathcal{F}^{(\ell)}_j  \rangle
= 0    \;, \;\;\;\ell= 1,2,\;\;\; j=1,2,3\;,
\end{equation}
%%%%%%%%%%%%%%%%%%%%%%%%%%%%%%%%
where $ \langle \cdot , \cdot \rangle$ denotes vector dot product. As we shall see later, (\ref{order-epsilon-eqn-Frdholm}) will lead to the six-wave interaction equations. Taking the vector inner product, we arrive at
%%%%%%%%%%%%%%%%%%%%%%%%%%%%%%%%%%%%%%%%%%%%%%%%%%%%%%%%%%%%%%%%%
         %%%%%%%%%%%%%%%%%%%%%%%%%%%%%%%%%
\begin{equation}
        \partial_{T}  A_j + \left( \mathbf{V}_{g,j} \cdot \nabla_{\mathbf{R}} \right) A_j
     =
       -   \frac{i\Omega^4 \omega_j (\omega_0^2 - \omega_j^2)}{\mu_0 c^2\left( (\omega_0^2 - \omega_j^2)^2 + \Omega^2\omega_0^2 \right)}
         \mathbb{P}^{(1)}_j \;,
          \label{Laser1}
         \end{equation}
%%%%%%%%%%%%%%%%%%%%%%%%%%%%%%%%
%%%%%%%%%%%%%%%%%%%%%%%%%%%%%%%%%
\begin{equation}
        \partial_{T}  B_j + \left( \mathbf{V}_{g,j} \cdot \nabla_{\mathbf{R}} \right) B_j
     =
          \frac{i\Omega^4 \omega_j (\omega_0^2 - \omega_j^2)}{\mu_0 c^2\left( (\omega_0^2 - \omega_j^2)^2 + \Omega^2\omega_0^2 \right)}
         \mathbb{P}^{(2)}_j \;,
          \label{Laser2}
         \end{equation}
         %%%%%%%%%%%%%%%%%%%%%%%%%%%%%%%%
where the group velocity is given by
%%%%%%%%%%%%%%%%%%%%
\begin{equation}
      \mathbf{V}_{g,j} = \frac{c^2\mathbf{k}_j}
     {\omega_j \left( 1 +   \frac{\Omega^{2}\omega_0^2}{(\omega_0^{2}-\omega_j^{2})^2)} \right)}\;,\;\;\; j=1,2,3\;.
        \end{equation}
 %%%%%%%%%%%%%%%%%%%%%%%
This group velocity coincides with the expression given by (\ref{group-velocity-j}) when $\hat{\chi}^{(L)}_{zz}(\omega_j)$ is given by (\ref{disp_E1}).
%%%%%%%%%%%%%%%%%%%%%%%%%%%%%%%%

%%%%%%%%%%%%%%%%%%%%%%%%%%%%%%%%%%%%%%%%%%%%%%%%%%%%%%%%%%%%%%%%
%%%%%%%%%%%%%%%%%%%%%%%%%%%%%%%%%%%%%%%%%%%%%%%%%%%%%%%%%%%%%%%%%%
\section{Inverse scattering transform for the 1+1D space-time shifted three-wave system}
\label{IST}
%%%%%%%%%%%%%%%%%%%%%%%%%%%%%%%%%%%%%%%%%%%%%%%%%%%%%%%%%%%%%%%%%%
In what follows we discuss the six-wave system in the form \eqref{six_wave_N_2} 
in $(1+1)$ dimensions with no $y$ dependence, i.e.
\begin{equation}
\label{six_wave_N}
        \partial_{t}N_{\ell j}(x,t) - \alpha_{\ell j}\partial_{x}N_{\ell j}(x,t) 
    =
    (\alpha_{\ell m}-\alpha_{mj})N_{\ell m}(x,t)N_{mj}(x,t)\,,
\end{equation}
which is an integrable system that arises as a compatibility condition 
between the $3\times3$ scattering problem
%%%%%%%%%%%%%%%%%%%%%%%%%%%%%%%%%
\begin{equation}
\label{scat}
    \partial_{x}v(x,t,k)=[ikD+N(x,t)]v(x,t,k)\,,
\end{equation}
%%%%%%%%%%%%%%%%%%%%%%%%%%%%%%%%%
where
%%%%%%%%%%%%%%%%%%%%%%%%%%%%%%%%%
\begin{equation}
    D=\begin{bmatrix}
        d_{1}&0&0\\
        0&d_{2}&0\\
        0&0&d_{3}
    \end{bmatrix},\;\;\;N(x,t)=\begin{bmatrix}
        0&N_{12}(x,t)&N_{13}(x,t)\\
        N_{21}(x,t)&0&N_{13}(x,t)\\
        N_{31}(x,t)&N_{32}(x,t)&0
    \end{bmatrix}\,,
\end{equation}
%%%%%%%%%%%%%%%%%%%%%%%%%%%%%%%%%
with $d_{j}=-C_{j}$ (note that $d_{1}>d_{2}>d_{3}$) and the time evolution equation
%%%%%%%%%%%%%%%%%%%%%%%%%%%%%%%%%
\begin{equation}
\label{time}
    \partial_{t}v(x,t,k)=\mathcal{T}v(x,t,k)\,,\;\;\;\mathcal{T} = \begin{bmatrix}
       -id_2d_3k & d_3N_{12}(x,t) & d_2N_{13}(x,t) \\ 
       d_3N_{21}(x,t) & -id_1d_3k & d_1N_{23}(x,t) \\
       d_2N_{31}(x,t) & d_1N_{32}(x,t) & -id_1d_2k \\
    \end{bmatrix}\,.
\end{equation}
%%%%%%%%%%%%%%%%%%%%%%%%%%%%%%%%%
By exploiting its connection to these two linear problems, the above six-wave system is solvable by the inverse scattering transform (IST). 
As mentioned earlier the IST for the classical three-wave system is well known. Space-time nonlocal variants of the three-wave system were considered by Gerdjikov et al in 2016 \cite{Gerd_3}; associated soliton solutions were obtained by dressing methods. Soliton solutions were also obtained using Darboux transformations in \cite{Sarfraz}.
The detailed  IST and associated bound states and soliton solutions necessary to solve the reverse space-time nonlocal three-wave system in comparison with the classical case were formulated in \cite{ALM_three}, whose notations and procedure we will follow in the present work.

%\mja{ NOTE: GERJIKOV ET AL (2016) DID NOT SOLVE THE PROBLEM BY IST. HE USES IST PART OF THE WAY AND THEN USES `DRESSING METHODS' TO FIND SOLITON SOLUTIONS. DRESSING METHODS ARE DRIECT METHODS OF SOLUTION; NOT IST SINCE THE DISCRETE SCATTERING DATA IS NOT FORMULATED VIA IST.} 

Here, we consider the nonlocal space-time shifted case \eqref{three_wave_shifted} which can be obtained by applying the symmetry reduction

%%%%%%%%%%%%%%%%%%%%%%%%%%%%%%%%%
\begin{equation}
\label{sym_N}
    N(x,t) = \Lambda N^{\dagger}(x_{0}-x,t_{0}-t)\Lambda\,,\;\;\; \Lambda=\begin{bmatrix}
     \epsilon_1 & 0 & 0 \\
       0 & \epsilon_2 & 0 \\
       0 & 0 & \epsilon_3
\end{bmatrix}\,,\;\;\;\epsilon_{1}\epsilon_{2}\epsilon_{3}=+1\,,
\end{equation}
%%%%%%%%%%%%%%%%%%%%%%%%%%%%%%%%%
(where $\dagger$ is the conjugate-transpose) to the matrix $N(x,t)$ followed by the rescaling in \eqref{scale_N_to_Q}. Note that applying \eqref{sym_N} followed by \eqref{scale_N_to_QR} is equivalent to applying \eqref{scale_N_to_QR} followed by \eqref{reduc_shifted}. Throughout this section, we give a brief overview of this solution process and give the relevant adaptations (particularly in the symmetries of the scattering data) to the case of the space-time shifted system. For a more detailed treatment, we refer the reader to \cite{ALM_three}.
%%%%%%%%%%%%%%%%%%%%%%%%%%%%%%%%%%%%%%%%%%%%%%%%%%%%%%%%%%%%%%%%%%
\subsection{Direct scattering problem}
%%%%%%%%%%%%%%%%%%%%%%%%%%%%%%%%%%%%%%%%%%%%%%%%%%%%%%%%%%%%%%%%%%
We assume that the potential matrix $N(x,t)$ decays to zero sufficiently rapidly as $x\rightarrow\pm\infty$, and define eigenfunctions of \eqref{scat} according to the following boundary conditions:
%%%%%%%%%%%%%%%%%%%%%%%%%%%%%%%%%
\bse
\begin{eqnarray}
  \phi_j(x,t,k)
    &\sim& e^{ikd_jx}\mathbf{\hat{e}}_{j}\,,\;\;\;\text{as}\;x\rightarrow-\infty\,,\\
      \psi_j(x,t,k)
    &\sim& e^{ikd_jx}\mathbf{\hat{e}}_{j}\,,\;\;\;\text{as}\;x\rightarrow+\infty\,,\;\;\;j=1,2,3\,.
\end{eqnarray}
\ese
%%%%%%%%%%%%%%%%%%%%%%%%%%%%%%%%%
Here and throughout the rest of the paper, $\mathbf{\hat{e}}_{1}=(1,0,0)^{T}$, $\mathbf{\hat{e}}_{2}=(0,1,0)^{T}$, and $\mathbf{\hat{e}}_{3}=(0,0,1)^{T}$. The two sets of eigenfunctions $\phi_{j}(x,t,k)$ and $\psi_{j}(x,t,k)$ each form a linearly independent set of solutions of the $3\times3$ system \eqref{scat}. As such, they can be expressed in terms of each other as follows: 
%%%%%%%%%%%%%%%%%%%%%%%%%%%%%%%%%
\bse
\begin{eqnarray}
     \phi_1(x,t,k) &=& a_{11}(t,k)\psi_1(x,t,k) + a_{12}(t,k)\psi_2(x,t,k) + a_{13}(t,k)\psi_3(x,t,k)\,,\\
    \phi_2(x,t,k) &=& a_{21}(t,k)\psi_1(x,t,k) + a_{22}(t,k)\psi_2(x,t,k) + a_{23}(t,k)\psi_3(x,t,k)\,, \\
    \phi_3(x,t,k) &=& a_{31}(t,k)\psi_1(x,t,k) + a_{32}(t,k)\psi_2(x,t,k) + a_{33}(t,k)\psi_3(x,t,k) \,,
\end{eqnarray}
\begin{eqnarray}
     \psi_1(x,t,k) &=& b_{11}(t,k)\phi_1(x,t,k) + b_{12}(t,k)\phi_2(x,t,k) + b_{13}(t,k)\phi_3(x,t,k)\,,\\
    \psi_2(x,t,k) &=& b_{21}(t,k)\phi_1(x,t,k) + b_{22}(t,k)\phi_2(x,t,k) + b_{23}(t,k)\phi_3(x,t,k) \,,\\
    \psi_3(x,t,k) &=& b_{31}(t,k)\phi_1(x,t,k) + b_{32}(t,k)\phi_2(x,t,k) + b_{33}(t,k)\phi_3(x,t,k)\,,
\end{eqnarray} 
\ese
%%%%%%%%%%%%%%%%%%%%%%%%%%%%%%%%%
where $a_{\ell j}(t,k)$ and $b_{\ell j}(t,k),\;\ell,j=1,2,3$ will be referred to as the scattering coefficients. The diagonal entries in $A=[a_{\ell j}]$ and $B=[b_{\ell j}]$ are the (inverse) transmission coefficients, and the off-diagonal entries determine the reflection coefficients
%%%%%%%%%%%%%%%%%%%%%%%%%%%%%%%%%
\begin{equation}
    \rho^{(a)}_{\ell j}(t,k)=\frac{a_{\ell j}(t,k)}{a_{\ell\ell}(t,k)},\;\;\;\rho^{(b)}_{\ell j}(t,k)=\frac{b_{\ell j}(t,k)}{b_{\ell\ell}(t,k)}\,,\;\;\;\ell\neq j\,.
\end{equation}
%%%%%%%%%%%%%%%%%%%%%%%%%%%%%%%%%
It will also be useful to define the following Jost functions which satisfy constant boundary conditions:
%%%%%%%%%%%%%%%%%%%%%%%%%%%%%%%%%
\begin{equation}
  M_j(x,t,k)= \phi_j (x,t,k)e^{-ikd_jx}\,,\;\;\;
   N_j(x,t,k)= \psi_j (x,t,k)e^{-ikd_jx}\;.
\end{equation} 
%%%%%%%%%%%%%%%%%%%%%%%%%%%%%%%%%
Indeed, it can be proven that $M_{3}(x,t,k)$ and $N_{1}(x,t,k)$ are analytic functions of the complex parameter $k$ in the upper-half plane, and as a consequence so are $a_{33}(t,k)$ and $b_{11}(t,k)$ while $M_{1}(x,t,k)$ and $N_{3}(x,t,k)$, along with $a_{11}(t,k)$ and $b_{33}(t,k),$ are analytic functions of $k$ in the lower-half plane. However, no such analyticity can be established for $M_{2}(x,t,k)$ or $N_{2}(x,t,k)$. To replace them with analytic eigenfunctions, a standard procedure is to define the adjoint (auxiliary) scattering problem 
%%%%%%%%%%%%%%%%%%%%%%%%%%%%%%%%%
\begin{equation}
\label{adj}
    \partial_{x}v^{\text{ad}}(x,t,k)=-[ikD+N^{T}(x,t)]v^{\text{ad}}(x,t,k)\,,
\end{equation}
%%%%%%%%%%%%%%%%%%%%%%%%%%%%%%%%%
whose solutions are related to solutions of the original scattering problem via $\partial_{x}(v^{T}v^{\text{ad}})=0$. In a similar fashion, we define the adjoint eigenfunctions
%%%%%%%%%%%%%%%%%%%%%%%%%%%%%%%%%
\bse
\begin{eqnarray}
  \phi^{\text{ad}}_j(x,t,k)
    &\sim& \epsilon_{j}e^{-ikd_jx}\mathbf{\hat{e}}_{j}\,,\;\;\;\text{as}\;x\rightarrow-\infty\,,\\
      \psi^{\text{ad}}_j(x,t,k)
    &\sim& \epsilon_{j}e^{-ikd_jx}\mathbf{\hat{e}}_{j}\,,\;\;\;\text{as}\;x\rightarrow+\infty\,,
\end{eqnarray}
\ese
%%%%%%%%%%%%%%%%%%%%%%%%%%%%%%%%%
and the corresponding adjoint Jost functions
%%%%%%%%%%%%%%%%%%%%%%%%%%%%%%%%%
\begin{equation}
  M^{\text{ad}}_j(x,t,k)= \phi^{\text{ad}}_j (x,t,k)e^{ikd_jx}\,,\;\;\;
   N^{\text{ad}}_j(x,t,k)= \psi^{\text{ad}}_j (x,t,k)e^{ikd_jx}\;.
\end{equation} 
%%%%%%%%%%%%%%%%%%%%%%%%%%%%%%%%%
It can again be proven that $M^{\text{ad}}_{3}(x,t,k)$ and $N^{\text{ad}}_{1}(x,t,k)$ are analytic in the upper-half $k$-plane, while $M^{\text{ad}}_{1}(x,t,k)$ and $N^{\text{ad}}_{3}(x,t,k)$ are analytic in the lower-half $k$-plane. Now, if $u^{\text{ad}}(x,t,k)$ and $w^{\text{ad}}(x,t,k)$ are two arbitrary solutions of \eqref{adj} then 
%%%%%%%%%%%%%%%%%%%%%%%%%%%%%%%%%
\begin{equation}
\label{adj_rel}
    v(x,t,k) = e^{ikdx}\big(u^{\text{ad}}(x,t,k) \times w^{\text{ad}}(x,t,k)\big)\,,
\end{equation} 
%%%%%%%%%%%%%%%%%%%%%%%%%%%%%%%%%
where $\times$ denotes the vector cross product and $d=d_1+d_2+d_3$, will be a solution of \eqref{scat}. So, we define the following two additional eigenfunctions of \eqref{scat},
%%%%%%%%%%%%%%%%%%%%%%%%%%%%%%%%%
\bse
\begin{eqnarray}
    \tau(x,t,k) &=& e^{ikdx}\big(\phi_1^{\text{ad}}(x,t,k) \times \psi_3^{\text{ad}}(x,t,k)\big)\,,\\
    \bar{\tau}(x,t,k) &=& e^{ikdx}\big(\phi_3^{\text{ad}}(x,t,k) \times \psi_1^{\text{ad}}(x,t,k)\big)\,,
\end{eqnarray} 
\ese
%%%%%%%%%%%%%%%%%%%%%%%%%%%%%%%%%
and the corresponding Jost functions 
%%%%%%%%%%%%%%%%%%%%%%%%%%%%%%%%%
\begin{equation}
      \chi(x,t,k)= \tau (x,t,k)e^{-ikd_2x}\,,\;\;\;
   \bar\chi(x,t,k)= \bar\tau (x,t,k)e^{-ikd_2x}\,,
\end{equation}
%%%%%%%%%%%%%%%%%%%%%%%%%%%%%%%%%
which are analytic functions of $k$ in the upper- and lower-half planes, respectively. Furthermore, by using \eqref{adj_rel} and comparing boundary conditions, one can establish the following useful representations for $\tau(x,t,k)$ and $\bar\tau(x,t,k)$:
%%%%%%%%%%%%%%%%%%%%%%%%%%%%%%%%%
\bse
\label{tau}
\begin{eqnarray}
    \tau(x,t,k)=b_{21}(t,k)\psi_1(x,t,k) - b_{11}(t,k)\psi_2(x,t,k)=a_{23}(t,k)\phi_3(x,t,k) - a_{33}(t,k)\phi_2(x,t,k)\,,\\
    \bar{\tau}(x,t,k)=b_{33}(t,k)\psi_2(x,t,k) - b_{23}(k)\psi_3(x,t,k) =a_{11}(t,k)\phi_2(x,t,k) - a_{21}(t,k)\phi_1(x,t,k)\,.  
\end{eqnarray}
\ese
Constructing the additional analytic structure of the scattering eigenfunctions from adjoint eigenfunctions is a convenient technique for determining fundamental analytic solutions of the $3\times3$ scattering problem \eqref{scat}, see e.g. \cite{Kaup1976, Prinari2006}. Alternative methods for constructing fundamental analytic solutions have been studied (see for instance \cite{Shabat,beals1984,Gerd_1}); they have
%\cite{Shabat,ZMNP,Gerd_1}), 
an advantage of constructing fundamental analytic solutions for suitable $n\times n$ Lax operators. However, the underlying discrete scattering data are more difficult to characterize. For the present 3×3 case, the use of the adjoint problem suffices.}

%straightforward generalization to $n\times n$ Lax operators. For the present $3\times3$ case, the use of the adjoint problem suffices.}
%%%%%%%%%%%%%%%%%%%%%%%%%%%%%%%%%
%%%%%%%%%%%%%%%%%%%%%%%%%%%%%%%%%%%%%%%%%%%%%%%%%%%%%%%
\subsection{Symmetries of the scattering data}
%%%%%%%%%%%%%%%%%%%%%%%%%%%%%%%%%%%%%%%%%%%%%%%%%%%%%%%%%%%%%%%%%%
It can be shown that the two scattering problems \eqref{scat} and \eqref{adj} with the reduction \eqref{sym_N} admit the following symmetry:
%%%%%%%%%%%%%%%%%%%%%%%%%%%%%%%%%
\begin{itemize}
    \item If $v(x,t,k)$ is a solution of the scattering problem \eqref{scat} with symmetry reduction \eqref{sym_N}, then $\Lambda v^{*}(x_{0}-x,t_{0}-t,-k^{*})$ is a solution of the adjoint scattering problem \eqref{adj}.
\end{itemize}
%%%%%%%%%%%%%%%%%%%%%%%%%%%%%%%%%
From this fact, we can derive space-time shifted symmetries connecting the eigenfunctions of both scattering problems. For instance, $\psi_{2}(x,t,k)$ and $\psi_{3}(x,t,k)$ are solutions of the direct problem, so $e^{-ikdx}(\psi_{2}(x,t,k)\times\psi_{3}(x,t,k))$ must be a solution of the adjoint problem, and it satisfies the same boundary condition as $\psi_{1}^{\text{ad}}(x,t,k)$, but without the factor of $\epsilon_{1}$. So, we have
%%%%%%%%%%%%%%%%%%%%%%%%%%%%%%%%%
\begin{equation}
\label{psi_1_ad}
    \epsilon_{1}\psi_{1}^{\text{ad}}(x,t,k)=e^{-ikdx}\big(\psi_{2}(x,t,k)\times\psi_{3}(x,t,k)\big)\,.
\end{equation}
%%%%%%%%%%%%%%%%%%%%%%%%%%%%%%%%%
Furthermore, taking into account the above symmetry then, $e^{ikdx}(\Lambda\psi_{2}^{*}(x_{0}-x,t_{0}-t,-k^{*})\times\Lambda\psi_{3}^{*}(x_{0}-x,t_{0}-t,-k^{*}))$ is a solution of the direct scattering problem. Then, comparing boundary conditions gives
%%%%%%%%%%%%%%%%%%%%%%%%%%%%%%%%%
\begin{equation}
    e^{ik(d_{2}+d_{3})x_{0}}\phi_{1}(x,t,k)=e^{ikdx}\big(\psi_{2}^{*}(x_{0}-x,t_{0}-t,-k^{*})\times\psi_{3}^{*}(x_{0}-x,t_{0}-t,-k^{*})\big)\,.
\end{equation}
%%%%%%%%%%%%%%%%%%%%%%%%%%%%%%%%%
Conjugating and letting $x\rightarrow x_{0}-x$, $t\rightarrow t_{0}-t$, $k\rightarrow-k^{*}$ gives
%%%%%%%%%%%%%%%%%%%%%%%%%%%%%%%%%
\begin{equation}
    e^{ik(d_{2}+d_{3})x_{0}}\phi_{1}^{*}\big(x_{0}-x,t_{0}-t,-k^{*})=e^{ikd(x_{0}-x)}(\psi_{2}(x,t,k)\times\psi_{3}(x,t,k)\big)\,.
\end{equation}
%%%%%%%%%%%%%%%%%%%%%%%%%%%%%%%%%
Comparing with \eqref{psi_1_ad} shows that
%%%%%%%%%%%%%%%%%%%%%%%%%%%%%%%%%
\begin{equation}
   \psi_{1}^{\text{ad}}(x,t,k)=\epsilon_{1}e^{-ikd_{1}x_{0}}\phi_{1}^{*}(x_{0}-x,t_{0}-t,-k^{*})\,.
\end{equation}
%%%%%%%%%%%%%%%%%%%%%%%%%%%%%%%%%
Following similar arguments, for $j=1,2,3$, one can get 
%%%%%%%%%%%%%%%%%%%%%%%%%%%%%%%%%
\bse
\label{sym_eig}
\begin{eqnarray}
    \psi_{j}^{\text{ad}}(x,t,k)&=&\epsilon_{j}e^{-ikd_{j}x_{0}}\phi_{j}^{*}(x_{0}-x,t_{0}-t,-k^{*})\,,\\
        \phi_{j}^{\text{ad}}(x,t,k)&=&\epsilon_{j}e^{-ikd_{j}x_{0}}\psi_{j}^{*}(x_{0}-x,t_{0}-t,-k^{*})\,.
\end{eqnarray}
\ese
%%%%%%%%%%%%%%%%%%%%%%%%%%%%%%%%%
In turn, corresponding symmetries of the scattering coefficients can be obtained. For example, we have
%%%%%%%%%%%%%%%%%%%%%%%%%%%%%%%%%
\bse
\begin{eqnarray}
\label{a1}
    a_{\ell 1}(t,k)&=&\frac{\det\big[\phi_{\ell}(x,t,k),\psi_{2}(x,t,k),\psi_{3}(x,t,k)\big]}{\det\big[\psi_{1}(x,t,k),\psi_{2}(x,t,k),\psi_{3}(x,t,k)\big]}\,,\\
\label{a2}
    a_{1\ell}(t,k)&=&\frac{\det\big[\psi^{\text{ad}}_{\ell}(x,t,k),\phi^{\text{ad}}_{2}(x,t,k),\phi^{\text{ad}}_{3}(x,t,k)\big]}{\det\big[\phi^{\text{ad}}_{1}(x,t,k),\phi^{\text{ad}}_{2}(x,t,k),\phi^{\text{ad}}_{3}(x,t,k)\big]}\,.
\end{eqnarray}
\ese
%%%%%%%%%%%%%%%%%%%%%%%%%%%%%%%%%%%%%%%%%%%%%%%%%%%%%%%%%%%%%%%%%%%%
Substituting the symmetries \eqref{sym_eig} into \eqref{a2} and comparing with \eqref{a1} gives
%%%%%%%%%%%%%%%%%%%%%%%%%%%%%%%%%%%%%%%%%%%%%%%%%%%%%%%%%%%%%%%%%%%%
\begin{eqnarray}
    a_{\ell1}(t,k)&=&\epsilon_{\ell}\epsilon_{1}e^{ik(d_{\ell}-d_{1})x_{0}}a^{*}_{1\ell}(t_{0}-t,-k^{*})\,.
\end{eqnarray}
By similar logic, for $\ell,j=1,2,3$, it can be shown that 
%%%%%%%%%%%%%%%%%%%%%%%%%%%%%%%%%%%%%%%%%%%%%%%%%%%%%%%%%%%%%%%%%%%%
\bse
\label{sym_scat}
\begin{eqnarray}
    a_{\ell j}(t,k)&=&\epsilon_{\ell}\epsilon_{j}e^{ik(d_{\ell}-d_{j})x_{0}}a^{*}_{j\ell}(t_{0}-t,-k^{*})\,,\\
    b_{\ell j}(t,k)&=&\epsilon_{\ell}\epsilon_{j}e^{ik(d_{\ell}-d_{j})x_{0}}b^{*}_{j\ell}(t_{0}-t,-k^{*})\,.
\end{eqnarray}
\ese
%%%%%%%%%%%%%%%%%%%%%%%%%%%%%%%%%
%%%%%%%%%%%%%%%%%%%%%%%%%%%%%%%%%%%%%%%%%%%%%%%%%%%%%%%%%%%%%%%%%%
\subsection{Time evolution and inverse scattering}
%%%%%%%%%%%%%%%%%%%%%%%%%%%%%%%%%%%%%%%%%%%%%%%%%%%%%%%%%%%%%%%%%%
The evolution equation \eqref{time} (in the limits $x\rightarrow\pm\infty$) can be used to determine the time dependence of the scattering coefficients. Since the situation is identical to \cite{ALM_three}, we simply state the results:
\bse
\begin{eqnarray}
\label{time_diag}
    &a_{\ell\ell}(t,k)=a_{\ell\ell}(0,k)\,,\;\;\;&b_{\ell\ell}(t,k)=b_{\ell\ell}(0,k)\,,\\
\label{time_off}
    &a_{\ell j}(t,k)=a_{\ell j}(0,k)e^{-i(d_{\ell}-d_{j})d_{m}kt}\,,\;\;\;&b_{\ell j}(t,k)=b_{\ell j}(0,k)e^{-i(d_{\ell}-d_{j})d_{m}kt}\,,
\end{eqnarray}
\ese
where in \eqref{time_off}, $\ell\neq j,\;j\neq m,\;m\neq\ell$ , $j,k,\ell=1,2,3$. 
%%%%%%%%%%%%%%%%%%%%%%%%%%%%%%%%%%%%%%%%%%%%%%%%%%%%%%%%%%%%%%%%%%
The inverse scattering problem associated with \eqref{scat} can be formulated as a Riemann-Hilbert problem in the upper/lower half $k$-plane with a jump across the real $k$ axis. For ease of presentation we will consider only the reflectionless case, where $\rho_{\ell j}^{(a)}=\rho_{\ell j}^{(b)}\equiv0$. The discrete eigenvalues are the zeros of the diagonal scattering coefficients $a_{11}(k)$, $a_{33}(k)$, $b_{11}(k)$, and $b_{33}(k)$, which are independent of time according to \eqref{time_diag}. We assume that $a_{33}(k)$ and  $b_{11}(k)$ have simple zeros $\alpha_j$, $j=1,\hdots,J$ and $\beta_n$, $n=1,\hdots,N$, respectively, in the upper-half plane; and that $b_{33}(k)$ and  $a_{11}(k)$ have simple zeros $\bar{\alpha}_j$, $j=1,\hdots,\bar{J}$ and $\bar{\beta}_n$, $n=1,\hdots,\bar{N}$, respectively, in the lower-half plane. Furthermore, for simplicity, we take all discrete eigenvalues to lie on the imaginary $k$ axis. Note that in general, due to the symmetry \eqref{sym_scat}, if $\alpha_{j}$ is a discrete eigenvalue, then so is $-\alpha_{j}^{*}$. In this simplified case, the solution of the Riemann-Hilbert problem reduces to the following linear system (again, we refer the reader to \cite{ALM_three} for details),
%%%%%%%%%%%%%%%%%%%%%%%%%%%%%%%%%%%%%%%%%%%%%%%%%%%%%%%%%%%%%%%%%%
\bse
\label{system}
\begin{eqnarray}
    N_1(x,t,k)&=&\mathbf{\hat{e}}_1 - \sum_{\ell=1}^{\Bar{N}}\frac{\Bar{\chi}(x,t,\Bar{\beta}_\ell)e^{i\Bar{\beta}_\ell(d_2-d_1)x}}{a_{21}(t,\Bar{\beta}_\ell)(k -\Bar{\beta}_\ell)\partial_{k}a_{11}(\Bar{\beta}_\ell)}\,,\\
    N_3(x,t,k)&=&\mathbf{\hat{e}}_3 + \sum_{m=1}^{J}\frac{\chi(x,t,\alpha_m)e^{i\alpha_m(d_2-d_3)x}}{a_{23}(t,\alpha_m)(k-\alpha_m)\partial_{k}a_{33}(\alpha_m)}\,,\\
\frac{\Bar{\chi}(x,t,k)}{b_{33}(t,k)}&=&\epsilon_{2}\mathbf{\hat{e}}_2 -\sum_{p=1}^{N}\frac{b_{21}(t,\beta_p)N_1(x,t,\beta_p)e^{i\beta_p(d_1-d_2)x}}{(k-\beta_p)\partial_{k}b_{11}(\beta_p)}\nonumber \\
&&-\,\sum_{q=1}^{\Bar{J}}\frac{b_{23}(t,\Bar{\alpha}_q)N_3(x,t,\Bar{\alpha}_q)e^{i\Bar{\alpha}_q(d_3-d_2)x}}{(k-\Bar{\alpha}_q)\partial_{k}b_{33}(\Bar{\alpha}_q)}\,,\\
\frac{\chi(x,t,k)}{b_{11}(t,k)}&=&-\epsilon_{2}\mathbf{\hat{e}}_2 +\sum_{p=1}^{N}\frac{b_{21}(t,\beta_p)N_1(x,t,\beta_p)e^{i\beta_p(d_1-d_2)x}}{(k-\beta_p)\partial_{k}b_{11}(\beta_p)}
\nonumber\\&&+\,\sum_{q=1}^{\Bar{J}}\frac{b_{23}(t,\Bar{\alpha}_q)N_3(x,t,\Bar{\alpha}_q)e^{i\Bar{\alpha}_q(d_3-d_2)x}}{(k-\Bar{\alpha}_q)\partial_{k}b_{33}(\Bar{\alpha}_q)}\,,
\end{eqnarray} 
\ese
%%%%%%%%%%%%%%%%%%%%%%%%%%%%%%%%%%%%%%%%%%%%%%%%%%%%%%%%%%%%%%%%%%
where $a_{21}(t,\bar\beta_{\ell})$, $a_{23}(t,\alpha_{m})$, $b_{21}(t,\beta_{p})$, and $b_{23}(t,\bar\alpha_{q})$ are so-called reduced normalization coefficients. Note that the off-diagonal scattering coefficients do not generically admit analytic continuation off the real $k$-axis, and the way we denote the reduced normalization coefficients is only for notational convenience. 

The Jost functions can be found from the above after solving the linear system with $J+\bar{J}+N+\bar{N}$ equations and unknowns obtained by evaluating \eqref{system} at the appropriate discrete eigenvalues. Finally, once the system above has been solved, the desired potentials $N_{12}(x,t)$, $N_{23}(x,t)$, and $N_{31}(x,t)$ which determine the solution of the space-time shifted three-wave system can be recovered by studying the asymptotic behavior of the Jost functions and comparing with \eqref{system}. Indeed, it can be proven that
%%%%%%%%%%%%%%%%%%%%%%%%%%%%%%%%%%%%%%%%%%%%%%%%%%%%%%%%%%%%%%%%%%
\bse
\label{recovery}
\begin{eqnarray}
    N_{21}(x,t) &\sim& ik(d_1-d_2)N_1^{(2)}(x,t,k)\,,\\
        N_{23}(x,t) &\sim& -ik(d_2-d_3)N_3^{(2)}(x,t,k)\,,\\
N_{31}(x,t) &\sim& ik(d_1-d_3)N_1^{(3)}(x,t,k)\,,
\end{eqnarray}
\ese
%%%%%%%%%%%%%%%%%%%%%%%%%%%%%%%%%%%%%%%%%%%%%%%%%%%%%%%%%%%%%%%%%%
as $|k|\rightarrow\infty$, where the superscripts on the right-hand side denote the vector component.
%%%%%%%%%%%%%%%%%%%%%%%%%%%%%%%%%%%%%%%%%%%%%%%%%%%%%%%%%%%%%%%%%%
\section{Soliton solutions}
%%%%%%%%%%%%%%%%%%%%%%%%%%%%%%%%%%%%%%%%%%%%%%%%%%%%%%%%%%%%%%%%%%
The 1-soliton solution can be obtained by setting $J=\bar{J}=N=\bar{N}=1$. In this case, \eqref{system} reduces to:
%%%%%%%%%%%%%%%%%%%%%%%%%%%%%%%%%%%%%%%%%%%%%%%%%%%%%%%%%%%%%%%%%%%%
\bse
\label{system_1}
\begin{eqnarray}
    N_1(x,t,k)&=&\mathbf{\hat{e}}_1 - \frac{\Bar{\chi}(x,t,\Bar{\beta}_1)e^{i\Bar{\beta}_1(d_2-d_1)x}}{a_{21}(t,\Bar{\beta}_1)(k -\Bar{\beta}_1)\partial_{k}a_{11}(\Bar{\beta}_1)}\,,\\
    N_3(x,t,k)&=&\mathbf{\hat{e}}_3 + \frac{\chi(x,t,\alpha_1)e^{i\alpha_1(d_2-d_3)x}}{a_{23}(t,\alpha_1)(k-\alpha_1)\partial_{k}a_{33}(\alpha_1)}\,,\\
\frac{\Bar{\chi}(x,t,k)}{b_{33}(t,k)}&=&\epsilon_{2}\mathbf{\hat{e}}_2 -\frac{b_{21}(t,\beta_1)N_1(x,t,\beta_1)e^{i\beta_1(d_1-d_2)x}}{(k-\beta_1)\partial_{k}b_{11}(\beta_1)}-\frac{b_{23}(t,\Bar{\alpha}_1)N_3(x,t,\Bar{\alpha}_1)e^{i\Bar{\alpha}_1(d_3-d_2)x}}{(k-\Bar{\alpha}_1)\partial_{k}b_{33}(\Bar{\alpha}_1)}\,,\\
\frac{\chi(x,t,k)}{b_{11}(t,k)}&=&-\epsilon_{2}\mathbf{\hat{e}}_2 +\frac{b_{21}(t,\beta_1)N_1(x,t,\beta_1)e^{i\beta_1(d_1-d_2)x}}{(k-\beta_1)\partial_{k}b_{11}(\beta_1)}
+\frac{b_{23}(t,\Bar{\alpha}_1)N_3(x,t,\Bar{\alpha}_1)e^{i\Bar{\alpha}_1(d_3-d_2)x}}{(k-\Bar{\alpha}_1)\partial_{k}b_{33}(\Bar{\alpha}_1)}\,.
\end{eqnarray} 
\ese
%%%%%%%%%%%%%%%%%%%%%%%%%%%%%%%%%%%%%%%%%%%%%%%%%%%%%%%%%%%%%%%%%%%%
It is straightforward to solve this system generically, but to specialize the result to the space-time shifted case, we need to reduce the degrees of freedom by expressing the scattering coefficients in terms of the discrete eigenvalues. First, we can express the diagonal scattering coefficients (and the necessary derivatives with respect to $k$) in terms of the eigenvalues by using established (see \cite{ALM_three}) trace formulae:
%%%%%%%%%%%%%%%%%%%%%%%%%%%%%%%%%%%%%%%%%%%%%%%%%%%%%%%%%%%%%%%%%%%%
\bse
\label{trace}
 \begin{eqnarray}
&\displaystyle a_{33}(k)=\frac{k-\alpha_1}{k-\Bar{\alpha}_1}\,,\;\;\;
    &b_{33}(k)= \frac{k-\Bar{\alpha}_1}{k-\alpha_1}\,,\\
    &\displaystyle b_{11}(k)= \frac{k-\beta_1}{k-\Bar{\beta}_1}\,,\;\;\;
    &a_{11}(k)= \frac{k-\Bar{\beta}_1}{k-\beta_1}\,.
\end{eqnarray}
\ese
%%%%%%%%%%%%%%%%%%%%%%%%%%%%%%%%%%%%%%%%%%%%%%%%%%%%%%%%%%%%%%%%%%%%
To find expressions for the reduced normalization coefficients, we need to employ additional symmetries. From \eqref{tau} we have
%%%%%%%%%%%%%%%%%%%%%%%%%%%%%%%%%%%%%%%%%%%%%%%%%%%%%%%%%%%%%%%%%%%%
\bse
\label{tau_disc}
\begin{eqnarray}
    &\tau(x,t,\alpha_1) = a_{23}(t,\alpha_1)\phi_3(x,t,\alpha_1)\,,\;\;\;
&\tau(x,t,\beta_1) = b_{21}(t,\beta_1)\psi_1(x,t,\beta_1)\,, \\
    &\Bar{\tau}(x,t,\Bar{\alpha}_1) = -b_{23}(t,\Bar{\alpha}_1)\psi_3(x,t,\Bar{\alpha}_1)\,,\;\;\;
    &\Bar{\tau}(x,t,\Bar{\beta}_1) = -a_{21}(t,\Bar{\beta}_1)\phi_1(x,t,\Bar{\beta}_1)\,.
\end{eqnarray}
\ese
%%%%%%%%%%%%%%%%%%%%%%%%%%%%%%%%%%%%%%%%%%%%%%%%%%%%%%%%%%%%%%%%%%%%
By applying the symmetries of the eigenfunctions \eqref{sym_eig}, it can be  deduced that if the eigenvalues are purely imaginary, then we have 
\bse
\label{sym_norm}
\begin{eqnarray}
      -\frac{\epsilon_3b^*_{21}(t,\beta_1)b_{21}(t_0-t,\beta_1)}{a_{33}(\beta_1)}e^{i\beta_1(d_1-d_2)x_0} = 1\,,\\
      -\frac{\epsilon_1b^*_{23}(t,\Bar{\alpha}_1)b_{23}(t_0-t,\Bar{\alpha}_1)}{a_{11}(\Bar{\alpha}_1)}e^{i\Bar{\alpha}_1(d_3-d_2)x_0} = 1\,,\\
      -\frac{\epsilon_3a^*_{21}(t,\Bar{\beta}_1)a_{21}(t_0-t,\Bar{\beta}_1)}{b_{33}(\Bar{\beta}_1)}e^{i\Bar{\beta}_1(d_1-d_2)x_0} = 1\,,\\
       -\frac{\epsilon_1a^*_{23}(t,\alpha_1)a_{23}(t_0-t,\alpha_1)}{b_{11}(\alpha_1)}e^{i\alpha_1(d_3-d_2)x_0} = 1\,.
\end{eqnarray}
\ese
%%%%%%%%%%%%%%%%%%%%%%%%%%%%%%%%%%%%%%%%%%%%%%%%%%%%%%%%%%%%%%%%%%%%
If we let $\alpha_{1}=iv_{1}$, $\beta_{1}=iv_{2}$, $\bar\alpha_{1}=-i\bar{v}_{1}$, $\bar\beta_{1}=-i\bar{v}_{2}$ with $v_{1},v_{2},\bar{v}_{1},\bar{v}_{2}\in\mathbb{R}^{+}$, recall that we defined $d_{j}=-C_{j}$ for $j=1,2,3$, and incorporate the trace formulae \eqref{trace} and the time evolution \eqref{time_off}, then from \eqref{sym_norm} we get
\bse
\label{red_norm}
%%%%%%%%%%%%%%%%%%%%%%%%%%%%%%%%%%%%%%%%%%%%%%%%%%%%%%%%%%%%%%%%%%%%
\begin{eqnarray}
    |b_{21}(0,iv_{2})|^2 &=& - \epsilon_3 \frac{v_2-v_1}{\Bar{v}_1+v_2}e^{v_2(C_2-C_1)(x_0-C_3t_0)}\,,\\
    |b_{23}(0,-i\bar{v}_{1})|^2 &=&  \epsilon_1 \frac{\Bar{v}_2-\Bar{v}_1}{\Bar{v}_1+v_2}e^{-\Bar{v}_1(C_2-C_3)(x_0-C_1t_0)}\,,\\
    |a_{21}(0,-i\bar{v}_{2})|^2 &=& - \epsilon_3 \frac{\Bar{v}_2-\Bar{v}_1}{v_1+\Bar{v}_2}e^{-\Bar{v}_2(C_2-C_1)(x_0-C_3t_0)}\,,\\
        |a_{23}(0,iv_{1})|^2 &=&  \epsilon_1 \frac{v_2-v_1}{v_1+\Bar{v}_2}e^{v_1(C_2-C_3)(x_0-C_1t_0)}\,.
\end{eqnarray} 
\ese
%%%%%%%%%%%%%%%%%%%%%%%%%%%%%%%%%%%%%%%%%%%%%%%%%%%%%%%%%%%%%%%%%%%%
Finally, the system \eqref{system_1} can be solved and the scattering data can be written fully in terms of the eigenvalues using \eqref{trace} and \eqref{red_norm}. Then, the solution of the space-time shifted three-wave system can be recovered using \eqref{recovery} followed by the rescaling in \eqref{scale_N_to_QR}. 

First, if we consider the case $\epsilon_{1}=1,\epsilon_{2}=-1,\epsilon_{3}=-1$, then $v_{2}>v_{1}$, $\bar{v}_{2}>\bar{v}_{1}$ and \eqref{red_norm} gives
\begin{eqnarray}
    b_{21}(0,iv_{2})&=&\sqrt{\frac{v_{2}-v_{1}}{\bar{v}_{1}+v_{2}}}e^{v_{2}(C_{2}-C_{1})(x_{0}-C_{3}t_{0})/2+i\theta_{1}}\,,\\
        b_{23}(0,-i\bar v_{1})&=&\sqrt{\frac{\bar v_{2}-\bar v_{1}}{\bar{v}_{1}+v_{2}}}e^{-\bar v_{1}(C_{2}-C_{3})(x_{0}-C_{1}t_{0})/2+i\theta_{2}}\,,\\
            a_{21}(0,-i\bar v_{2})&=&\sqrt{\frac{\bar v_{2}-\bar v_{1}}{{v}_{1}+\bar v_{2}}}e^{-\bar v_{2}(C_{2}-C_{1})(x_{0}-C_{3}t_{0})/2+i\theta_{3}}\,,\\
                    a_{23}(0,i v_{1})&=&\sqrt{\frac{ v_{2}- v_{1}}{{v}_{1}+\bar v_{2}}}e^{ v_{1}(C_{2}-C_{3})(x_{0}-C_{1}t_{0})/2+i\theta_{4}}\,,
\end{eqnarray}
where the $\theta_{j}$ are arbitrary phases. After simplification, one can obtain the solution
\bse
\begin{eqnarray}
    &\displaystyle Q_{1}(x,t)=\frac{iG_{1}(v_{1}+\bar{v}_{1})}{D(x,t)}\sqrt{\frac{v_{2}-v_{1}}{v_{1}+\bar{v}_{2}}}e^{\xi_{1}(x,t)}\Bigg[1+\sqrt{\frac{(\bar{v}_{2}+v_{1})(\bar{v}_{2}-\bar{v}_{1})}{(v_{2}-v_{1})(\bar{v}_{1}+v_{2})}}e^{\xi_{2}(x,t)+\bar{\xi}_{2}(x,t)}\Bigg]\,,\\
    &\displaystyle Q_{2}(x,t)=\frac{iG_{2}}{D(x,t)}\frac{(\bar{v}_{1}+v_{1})(v_{2}+\bar{v}_{2})}{\sqrt{(\bar{v}_{1}+v_{2})(v_{1}+\bar{v}_{2})}}e^{\bar{\xi}_{1}(x,t)+\bar{\xi}_{2}(x,t)}\,,\\
    &\displaystyle Q_{3}(x,t)=\frac{iG_{3}(v_{2}+\bar{v}_{2})}{D^{*}(x_{0}-x,t_{0}-t)}\sqrt{\frac{\bar{v}_{2}-\bar{v}_{1}}{v_{1}+\bar{v}_{2}}}e^{-\bar{\xi}_{2}(x,t)}\Bigg[1+\sqrt{\frac{(\bar{v}_{2}+v_{1})({v}_{2}-{v}_{1})}{(\bar v_{2}-\bar v_{1})(\bar{v}_{1}+v_{2})}}e^{-\xi_{1}(x,t)-\bar{\xi}_{1}(x,t)}\Bigg]\,,
\end{eqnarray}
\ese
where 
\begin{eqnarray}
    D(x,t)&=&\Bigg[1-\sqrt{\frac{(v_{2}-v_{1})(\bar{v}_{2}-\bar{v}_{1})}{(\bar{v}_{1}+v_{2})(v_{1}+\bar{v}_{2})}}e^{\xi_{2}(x,t)+\bar{\xi}_{2}(x,t)}\Bigg]\Bigg[1+\sqrt{\frac{(v_{2}-v_{1})(\bar{v}_{2}-\bar{v}_{1})}{(\bar{v}_{1}+v_{2})(v_{1}+\bar{v}_{2})}}e^{\xi_{1}(x,t)+\bar{\xi}_{1}(x,t)}\Bigg]\nonumber\\
    &&-\,\frac{(\bar{v}_{1}+v_{1})(\bar{v}_{2}+v_{2})}{(\bar{v}_{1}+v_{2})(v_{1}+\bar{v}_{2})}e^{\xi_{1}(x,t)+\bar\xi_{1}(x,t)+\xi_{2}(x,t)+\bar\xi_{2}(x,t)}\,,
\end{eqnarray}
and
\bse
\begin{eqnarray}
    \xi_{1}(x,t)&=&-i\theta_{4}+v_{1}(C_{2}-C_{3})[(x-C_{1}t)-(x_{0}-C_{1}t_{0})/2]\,,\\
    \bar\xi_{1}(x,t)&=&i\theta_{2}+\bar v_{1}(C_{2}-C_{3})[(x-C_{1}t)-(x_{0}-C_{1}t_{0})/2]\,,\\
    \xi_{2}(x,t)&=&i\theta_{1}+v_{2}(C_{1}-C_{2})[(x-C_{3}t)-(x_{0}-C_{3}t_{0})/2]\,,\\
    \bar\xi_{2}(x,t)&=&-i\theta_{3}+v_{2}(C_{1}-C_{2})[(x-C_{3}t)-(x_{0}-C_{3}t_{0})/2]\,,\\
    G_{1}&=&(C_{2}-C_{3})\sqrt{(C_{2}-C_{1})(C_{3}-C_{1})}\,,\\G_{2}&=&(C_{3}-C_{1})\sqrt{(C_{2}-C_{1})(C_{3}-C_{2})}\,,\\G_{3}&=&(C_{1}-C_{2})\sqrt{(C_{3}-C_{1})(C_{3}-C_{2})}\,.
\end{eqnarray}
\ese
This solution is guaranteed to be non-singular if we choose the phases such that $\theta_{1}-\theta_{3}=\pi+2k_{1}\pi$, $\theta_{2}-\theta_{4}=2k_{2}\pi$, for $k_{1},k_{2}\in\mathbb{Z}$.

\begin{figure}[ht]
\centering
\includegraphics[width=\textwidth]{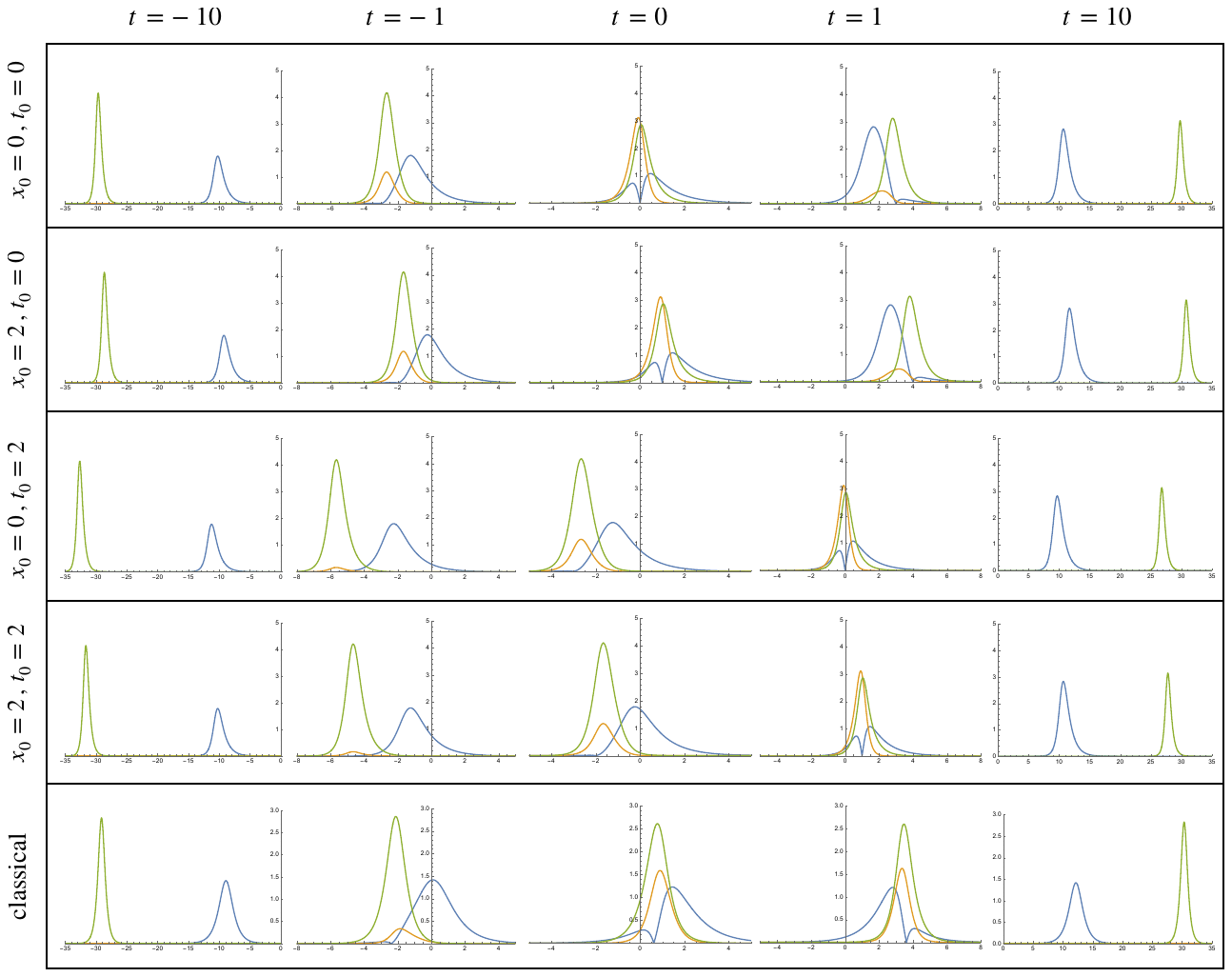}
\caption{Soliton solutions $|Q_{1}|$ (blue), $|Q_{2}|$ (yellow), and $|Q_{3}|$ (green) corresponding to the parameters $\epsilon_{1}=1$, $\epsilon_{2}=-1$, $\epsilon_{3}=-1$, $v_{1}=1$, $\bar{v}_{1}=2$, $v_{2}=2$, $\bar{v}_{2}=3$, $C_{1}=1$, $C_{2}=2$, $C_{3}=3$, $\theta_{1}=\pi$, $\theta_{2}=\theta_{3}=\theta_{4}=0$, with no shift (row 1), a spatial shift (row 2), a temporal shift (row 3), and a spatial and temporal shift (row 4), plotted for various fixed times (columns). For comparison, the fifth row displays an example of a soliton solution of the classical three-wave system (see \cite{ALM_three}) with the same group velocities $C_{1}=1$, $C_{2}=2$, $C_{3}=3$.}
\label{fig1}
\end{figure}

On the other hand, if $\epsilon_{1}=-1$, $\epsilon_{2}=-1$, $\epsilon_{3}=1$, then $v_{2}<v_{1}$, $\bar{v}_{2}<\bar{v}_{1}$ and \eqref{red_norm} gives
\begin{eqnarray}
    b_{21}(iv_{2},0)&=&\sqrt{\frac{v_{1}-v_{2}}{\bar{v}_{1}+v_{2}}}e^{v_{2}(C_{2}-C_{1})(x_{0}-C_{3}t_{0})/2+i\theta_{1}}\,,\\
        b_{23}(-i\bar v_{1},0)&=&\sqrt{\frac{\bar v_{1}-\bar v_{2}}{\bar{v}_{1}+v_{2}}}e^{-\bar v_{1}(C_{2}-C_{3})(x_{0}-C_{1}t_{0})/2+i\theta_{2}}\,,\\
            a_{21}(-i\bar v_{2},0)&=&\sqrt{\frac{\bar v_{1}-\bar v_{2}}{{v}_{1}+\bar v_{2}}}e^{-\bar v_{2}(C_{2}-C_{1})(x_{0}-C_{3}t_{0})/2+i\theta_{3}}\,,\\
                    a_{23}(i v_{1},0)&=&\sqrt{\frac{ v_{1}- v_{2}}{{v}_{1}+\bar v_{2}}}e^{ v_{1}(C_{2}-C_{3})(x_{0}-C_{1}t_{0})/2+i\theta_{4}}\,.
\end{eqnarray}
The solution in this case is
\begin{eqnarray}
    &\displaystyle Q_{1}(x,t)=\frac{iG_{1}(v_{1}+\bar{v}_{1})}{D(x,t)}\sqrt{\frac{v_{1}-v_{2}}{v_{1}+\bar{v}_{2}}}e^{\xi_{1}(x,t)}\Bigg[1+\sqrt{\frac{(\bar{v}_{2}+v_{1})(\bar{v}_{1}-\bar{v}_{2})}{(v_{1}-v_{2})(\bar{v}_{1}+v_{2})}}e^{\xi_{2}(x,t)+\bar{\xi}_{2}(x,t)}\Bigg]\,,\\
    &\displaystyle Q_{2}(x,t)=\frac{iG_{2}}{D(x,t)}\frac{(\bar{v}_{1}+v_{1})(v_{2}+\bar{v}_{2})}{\sqrt{(\bar{v}_{1}+v_{2})(v_{1}+\bar{v}_{2})}}e^{\bar{\xi}_{1}(x,t)+\bar{\xi}_{2}(x,t)}\,,\\
    &\displaystyle Q_{3}(x,t)=\frac{iG_{3}(v_{2}+\bar{v}_{2})}{D^{*}(x_{0}-x,t_{0}-t)}\sqrt{\frac{\bar{v}_{1}-\bar{v}_{2}}{v_{1}+\bar{v}_{2}}}e^{-\bar{\xi}_{2}(x,t)}\Bigg[1+\sqrt{\frac{(\bar{v}_{2}+v_{1})({v}_{1}-{v}_{2})}{(\bar v_{1}-\bar v_{2})(\bar{v}_{1}+v_{2})}}e^{-\xi_{1}(x,t)-\bar{\xi}_{1}(x,t)}\Bigg]\,,
\end{eqnarray}
where
\begin{eqnarray}
    D(x,t)&=&\Bigg[1+\sqrt{\frac{(v_{1}-v_{2})(\bar{v}_{1}-\bar{v}_{2})}{(\bar{v}_{1}+v_{2})(v_{1}+\bar{v}_{2})}}e^{\xi_{2}(x,t)+\bar{\xi}_{2}(x,t)}\Bigg]\Bigg[1-\sqrt{\frac{(v_{1}-v_{2})(\bar{v}_{1}-\bar{v}_{2})}{(\bar{v}_{1}+v_{2})(v_{1}+\bar{v}_{2})}}e^{\xi_{1}(x,t)+\bar{\xi}_{1}(x,t)}\Bigg]\nonumber\\
    &&-\,\frac{(\bar{v}_{1}+v_{1})(\bar{v}_{2}+v_{2})}{(\bar{v}_{1}+v_{2})(v_{1}+\bar{v}_{2})}e^{\xi_{1}(x,t)+\bar\xi_{1}(x,t)+\xi_{2}(x,t)+\bar\xi_{2}(x,t)}\,,
\end{eqnarray}
%and the
 with the other definitions the same as above.  
Figure \ref{fig1} shows examples of soliton solutions in the case of no shift (first row), a spatial shift (second row), a temporal shift (third row), and a spatial and temporal shift (fourth row). For comparison, the fifth row shows a typical example of the soliton solution of the classical three-wave interaction system \eqref{three_wave_classical} (see \cite{ALM_three} for the explicit form of the classical solution). Each column is a snapshot in time. From the first column and the final column, observe that the amplitudes of $Q_{1}$ and $Q_{3}$ generically change after they interact with each other and with $Q_{2}$ in all nonlocal cases. The amount by which they change depends only on the discrete eigenvalues, and in the special case where $v_{1}=\bar{v}_{1},\,v_{2}=\bar{v}_{2}$, the amplitudes remain unchanged. In contrast, the amplitudes are always preserved in the local case, as can be seen in the fifth row. Additionally, by comparing the second, third, and fourth columns of the first four rows, one can see the spatial shifting and time delay effects of the parameters $x_{0}$ and $t_{0}$. It is also seen that the $x_0,t_0$ shifts do not affect the final amplitudes.

We remark that the IST for the full 2+1 dimensional system \eqref{six_wave_QR} and its reductions has been studied, though the analysis is more complicated than the 1+1 dimensional case. Solutions of the classical three-wave system in two dimensions are known (see for example \cite{Kaup1981,Gerd_2}), but to the best of our knowledge, no solutions of its reverse space-time or shifted nonlocal variants have been presented. This is outside the scope of this paper.
%%%%%%%%%%%%%%%%%%%%%%%%%%%%%%%%%%%%%%%%%%%%%%%%%%%%%%%%%%%%%%%%%%%%%
\section{Conservation laws and Hamiltonian structure}
%%%%%%%%%%%%%%%%%%%%%%%%%%%%%%%%%%%%%%%%%%%%%%%%%%%%%%%%%%%%%%%%%%%%%
In \cite{Hab}, an infinite set of conservation laws (generated by a recurrence) for the general $N\times N$ version of the scattering problem \eqref{scat} associated with the six-wave interaction system was derived. In this section, we follow  
that work but specialize the result to both the classical and space-time shifted nonlocal three-wave systems. Finally, we connect some of the conserved quantities that we obtain from the recurrence to the infinite-dimensional Hamiltonian structure of the six-wave system. Throughout this section, we suppress $(x,t)$ dependence when it is not crucial.
%%%%%%%%%%%%%%%%%%%%%%%%%%%%%%%%%%%%%%%%%%%%%%%%%%%%%%%%%%%%%
\subsection{Infinite set of conservation laws}
%%%%%%%%%%%%%%%%%%%%%%%%%%%%%%%%%%%%%%%%%%%%%%%%%%%%%%%%%%%%%
Consider a fundamental analytic matrix solution $\Phi=[\Phi_{\ell j}]$ of \eqref{scat} that has the large-$k$ asymptotic behavior
\begin{equation}
    \Phi\sim \begin{bmatrix}
        a_{1}e^{ikd_{1}x}&0&0\\
        0&a_{2}e^{ikd_{2}x}&0\\
        0&0&a_{3}e^{ikd_{3}x}
    \end{bmatrix}+\mathcal{O}(1/k)\,,\;\;\;\text{as}\;|k|\rightarrow\infty\,,
\end{equation}
where the $a_{j}$ are independent of $k$. For example, we could take $\Phi=(\psi_{1},\tau,\phi_{3})$ in the upper-half $k$ plane. Define
\begin{equation}
    \Gamma_{\ell j}\equiv\frac{\Phi_{\ell j}}{\Phi_{jj}}\,,\;\ell\neq j\,,\;\;\;\Gamma_{jj}\equiv1\,,
\end{equation}
noting that $\Gamma_{\ell j}\sim\delta_{\ell j}+\mathcal{O}(1/k)$ as $|k|\rightarrow\infty$.
Now, starting from
\begin{equation}
    \partial_{x}\left[\partial_{t}\log\Phi_{jj}\right]=\partial_{t}\left[\partial_{x}\log\Phi_{jj}\right]\implies
    \partial_{x}\left[\frac{\partial_{t}\Phi_{jj}}{\Phi_{jj}}\right]=\partial_{t}\left[\frac{\partial_{x}\Phi_{jj}}{\Phi_{jj}}\right]\,,
\end{equation}
and applying \eqref{scat} and \eqref{time} we get
\begin{equation}
\label{compat}
        \partial_{x}\sum_{\ell\neq j}\alpha_{j\ell}N_{j\ell}\Gamma_{\ell j}=\partial_{t}\sum_{\ell\neq j}N_{j\ell}\Gamma_{\ell j}\,,
\end{equation}
with $\alpha_{j\ell}$ defined as below \eqref{six_wave_N}. Based on its large-$k$ asymptotic behavior, assume a Laurent series expansion for $\Gamma_{\ell j}$ with $\ell\neq j$ in the form
\begin{equation}
\label{laurent}
    \Gamma_{\ell j}=\sum_{p=1}^{\infty}\frac{\Gamma_{\ell j}^{(p)}}{k^{p}}\,.
\end{equation}
Upon substituting \eqref{laurent} and equating coefficients of $1/k^{p}$, \eqref{compat} gives an infinite set of local conservation laws:
\begin{equation}
    \partial_{x}\sum_{\ell\neq j}\alpha_{j\ell}N_{j\ell}\Gamma_{\ell j}^{(p)}=\partial_{t}\sum_{\ell\neq j}N_{j\ell}\Gamma_{\ell j}^{(p)}\,,\;\;\;j=1,2,3\,,\;\;\;p=1,\hdots,\infty\,,
\end{equation}
which can be integrated to obtain the global conserved quantities:
\begin{equation}
\label{quantities}
    \mathcal{C}_{j}^{(p)}=\sum_{\ell\neq j}\int_{\mathbb{R}}N_{j\ell}(x,t)\Gamma_{\ell j}^{(p)}(x,t)\,dx\,,\;\;\;j=1,2,3\,,\;\;\;p=1,\hdots,\infty\,.
\end{equation}
Furthermore, directly from \eqref{scat}, it can be found that $\Gamma_{\ell j}$ satisfies the Riccati-type differential equation
\begin{equation}
\label{riccati}
    ik(d_{\ell}-d_{j})\Gamma_{\ell j}=\partial_{x}\Gamma_{\ell j}-N_{\ell m}\Gamma_{mj}+\Gamma_{\ell j}N_{jm}\Gamma_{mj}\,,
\end{equation}
with $m\neq j$ and $m\neq \ell$. Substituting the Laurent series, with the additional $p=0$ term corresponding to $\Gamma_{\ell j}^{(0)}\equiv\delta_{\ell j}$, into \eqref{riccati} gives
\begin{equation}
    ik(d_{\ell}-d_{j})\sum_{p=0}^{\infty}\frac{\Gamma_{\ell j}^{(p)}}{k^{p}}=\partial_{x}\sum_{p=0}^{\infty}\frac{\Gamma_{\ell j}^{(p)}}{k^{p}}-N_{\ell m}\sum_{p=0}^{\infty}\frac{\Gamma_{mj}^{(p)}}{k^{p}}+\sum_{p=0}^{\infty}\frac{\Gamma_{\ell j}^{(p)}}{k^{p}}N_{jm}\sum_{s=0}^{\infty}\frac{\Gamma_{mj}^{(s)}}{k^{s}}\,.
\end{equation}
The $\mathcal{O}(k)$ terms automatically cancel, and the $\mathcal{O}(1)$ terms give
\begin{equation}
\label{gamma_1}
        \Gamma_{\ell j}^{(1)}=\frac{iN_{\ell j}}{d_{\ell}-d_{j}}\,,\;\;\;\ell\neq j\,.
\end{equation}
The $\mathcal{O}(1/k)$ terms give
\begin{equation}
\label{gamma_2}
    \Gamma_{\ell j}^{(2)}=\frac{1}{(d_{\ell}-d_{j})^{2}}\partial_{x}N_{\ell j}-\frac{1}{(d_{\ell}-d_{j})(d_{m}-d_{j})}N_{\ell m}N_{mj}\,,\;\;\;\ell\neq j,\;j\neq m,\;m\neq\ell\,.
\end{equation}
All subsequent terms can be determined by a recurrence. In particular, after rewriting the double sum, for $\ell \neq j$ we have
\begin{equation}
    i(d_{\ell}-d_{j})\sum_{p=0}^{\infty}\frac{\Gamma_{\ell j}^{(p+1)}}{k^{p}}=\partial_{x}\sum_{p=1}^{\infty}\frac{\Gamma_{\ell j}^{(p)}}{k^{p}}-N_{\ell m}\sum_{p=1}^{\infty}\frac{\Gamma_{mj}^{(p)}}{k^{p}}+\sum_{p=1}^{\infty}\sum_{s=1}^{p-1}N_{jm}\frac{\Gamma_{\ell j}^{(p+s)}\Gamma_{mj}^{(s)}}{k^{p}}\,.
\end{equation}
Equating coefficients of $1/k^{p}$ for $p\geq2$ gives
\begin{equation}
  i(d_{\ell}-d_{j})\Gamma_{\ell j}^{(p+1)}=\partial_{x}\Gamma_{\ell j}^{(p)}-N_{\ell m}\Gamma_{mj}^{(p)}+\sum_{s=1}^{p-1}N_{jm}\Gamma_{\ell j}^{(p+s)}\Gamma_{mj}^{(s)}\,.
\end{equation}
%%%%%%%%%%%%%%%%%%%%%%%%%%%%%%%%%%%%%%%%%%%%%%%%%%%%%
\subsection{The first few conserved quantities}
%%%%%%%%%%%%%%%%%%%%%%%%%%%%%%%%%%%%%%%%%%%%%%%%%%%%%
From \eqref{quantities} and \eqref{gamma_1}, the quantities
\begin{equation}
    \sum_{\ell\neq j}\int_{\mathbb{R}}N_{j\ell}\Gamma_{\ell j}^{(1)}\,dx=
\sum_{\ell\neq j}\frac{i}{d_{\ell}-d_{j}}\int_{\mathbb{R}}N_{j\ell}N_{\ell j}\,dx\,,
\end{equation}
are conserved. Setting $j=1,2,3$, performing the rescaling in \eqref{scale_N_to_QR}, and putting $d_{j}=-C_{j}$, the first three conserved quantities for the generic six-wave system are (proportional to):
\bse
\label{old_quantities}
\begin{eqnarray}
    \mathcal{C}_{1}^{(1)}&=&\int_{\mathbb{R}}Q_{2}R_{2}\,dx-\int_{\mathbb{R}}Q_{3}R_{3}\,dx\,,\\
    \mathcal{C}_{2}^{(1)}&=&\int_{\mathbb{R}}Q_{3}R_{3}\,dx-\int_{\mathbb{R}}Q_{1}R_{1}\,dx\,,\\
    \mathcal{C}_{3}^{(1)}&=&\int_{\mathbb{R}}Q_{1}R_{1}\,dx-\int_{\mathbb{R}}Q_{2}R_{2}\,dx\,.
\end{eqnarray}
\ese
These conserved quantities are well-known, and be also be derived directly from the six-wave system. Note that any one of them could be written as a linear combination of the other two. If we incorporate the classical symmetry reduction \eqref{reduc_classical}, then these can be written as
\bse
\begin{eqnarray}
    \mathcal{C}_{1}^{(1)}&=&\tilde\epsilon_{2}\int_{\mathbb{R}}Q_{2}(x,t)Q^{*}_{2}(x,t)\,dx-\tilde\epsilon_{3}\int_{\mathbb{R}}Q_{3}(x,t)Q^{*}_{3}(x,t)\,dx\,,\\
    \mathcal{C}_{2}^{(1)}&=&\tilde\epsilon_{3}\int_{\mathbb{R}}Q_{3}(x,t)Q^{*}_{3}(x,t)\,dx-\tilde\epsilon_{2}\int_{\mathbb{R}}Q_{1}(x,t)Q^{*}_{1}(x,t)\,dx\,,\\
    \mathcal{C}_{3}^{(1)}&=&\tilde\epsilon_{1}\int_{\mathbb{R}}Q_{1}(x,t)Q^{*}_{1}(x,t)\,dx-\tilde\epsilon_{2}\int_{\mathbb{R}}Q_{2}(x,t)Q^{*}_{2}(x,t)\,dx\,.
\end{eqnarray}
\ese
On the other hand, if we apply the shifted symmetry reduction \eqref{reduc_shifted}, then they become
\bse
\begin{eqnarray}
    \mathcal{C}_{1}^{(1)}&=&\epsilon_{2}\int_{\mathbb{R}}Q_{2}(x,t)Q_{2}^{*}(x_{0}-x,t_{0}-t)\,dx+\epsilon_{3}\int_{\mathbb{R}}Q_{3}(x,t)Q_{3}^{*}(x_{0}-x,t_{0}-t)\,dx\,,\\
    \mathcal{C}_{2}^{(1)}&=&\epsilon_{3}\int_{\mathbb{R}}Q_{3}(x,t)Q_{3}^{*}(x_{0}-x,t_{0}-t)\,dx-\epsilon_{1}\int_{\mathbb{R}}Q_{1}(x,t)Q_{1}^{*}(x_{0}-x,t_{0}-t)\,dx\,,\\
    \mathcal{C}_{3}^{(1)}&=&\epsilon_{1}\int_{\mathbb{R}}Q_{1}(x,t)Q_{1}^{*}(x_{0}-x,t_{0}-t)\,dx+\epsilon_{2}\int_{\mathbb{R}}Q_{2}(x,t)Q_{2}^{*}(x_{0}-x,t_{0}-t)\,dx\,.
\end{eqnarray}
\ese
From \eqref{quantities} and \eqref{gamma_2}, the second set of conserved quantities is
\begin{equation}
    \sum_{\ell\neq j}\int_{\mathbb{R}}N_{j\ell}\Bigg\{\frac{1}{(d_{\ell}-d_{j})^{2}}\partial_{x}N_{\ell j}-\frac{1}{(d_{\ell}-d_{j})(d_{m}-d_{j})}N_{\ell m}N_{mj}\Bigg\}\,dx\,.
\end{equation}
with $j=1,2,3$, $\ell\neq j$, and $m\neq\ell,j$. After simplification and applying the rescaling \eqref{scale_N_to_QR}, we find that these three quantities are (proportional to):
\bse
\label{new_quantities}
\begin{eqnarray}
    \mathcal{C}_{1}^{(2)}&=&\int_{\mathbb{R}}\bigg[(C_{1}-C_{2})R_{2}\partial_{x}Q_{2}+(C_{1}-C_{3})R_{3}\partial_{x}Q_{3}-iQ_{1}Q_{2}Q_{3}-iR_{1}R_{2}R_{3}\bigg]\,dx\,,\\
    \mathcal{C}_{2}^{(2)}&=&\int_{\mathbb{R}}\bigg[(C_{2}-C_{3})R_{3}\partial_{x}Q_{3}+(C_{2}-C_{1})R_{1}\partial_{x}Q_{1}-iQ_{1}Q_{2}Q_{3}-iR_{1}R_{2}R_{3}\bigg]\,dx\,,\\
    \mathcal{C}_{3}^{(2)}&=&\int_{\mathbb{R}}\bigg[(C_{3}-C_{1})R_{1}\partial_{x}Q_{1}+(C_{3}-C_{2})R_{2}\partial_{x}Q_{2}-iQ_{1}Q_{2}Q_{3}-iR_{1}R_{2}R_{3}\bigg]\,dx\,.
\end{eqnarray}
\ese
Applying the classical symmetry reduction \eqref{reduc_classical} gives
\bse
\begin{eqnarray}
    \mathcal{C}_{1}^{(2)}&=&\int_{\mathbb{R}}\bigg[\tilde\epsilon_{2}(C_{1}-C_{2})Q^{*}_{2}\partial_{x}Q_{2}+\tilde\epsilon_{3}(C_{1}-C_{3})Q^{*}_{3}\partial_{x}Q_{3}\bigg]\,dx+2i\,\text{Re}\int_{\mathbb{R}}Q_{1}Q_{2}Q_{3}\,dx\,,\\
    \mathcal{C}_{2}^{(2)}&=&\int_{\mathbb{R}}\bigg[\tilde\epsilon_{3}(C_{2}-C_{3})Q^{*}_{3}\partial_{x}Q_{3}+\tilde\epsilon_{1}(C_{2}-C_{1})Q^{*}_{1}\partial_{x}Q_{1}\bigg]\,dx+2i\,\text{Re}\int_{\mathbb{R}}Q_{1}Q_{2}Q_{3}\,dx\,,\\
    \mathcal{C}_{3}^{(2)}&=&\int_{\mathbb{R}}\bigg[\tilde\epsilon_{1}(C_{3}-C_{1})Q^{*}_{1}\partial_{x}Q_{1}+\tilde\epsilon_{2}(C_{3}-C_{2})Q^{*}_{2}\partial_{x}Q_{2}\bigg]\,dx+2i\,\text{Re}\int_{\mathbb{R}}Q_{1}Q_{2}Q_{3}\,dx\,.
\end{eqnarray}
\ese
Applying the shifted symmetry reduction \eqref{reduc_shifted} gives
\bse
\begin{eqnarray}
    \mathcal{C}_{1}^{(2)}&=&\int_{\mathbb{R}}\bigg[-\epsilon_{2}(C_{1}-C_{2})Q^{*}_{2}(x_{0}-x,t_{0}-t)\partial_{x}Q_{2}(x,t)\nonumber\\&&\qquad+\,\epsilon_{3}(C_{1}-C_{3})Q^{*}_{3}(x_{0}-x,t_{0}-t)\partial_{x}Q_{3}(x,t)\bigg]\,dx-i\Upsilon\,,\\
    \mathcal{C}_{2}^{(2)}&=&\int_{\mathbb{R}}\bigg[\epsilon_{3}(C_{2}-C_{3})Q^{*}_{3}(x_{0}-x,t_{0}-t)\partial_{x}Q_{3}(x,t)\nonumber\\&&\qquad+\,\epsilon_{1}(C_{2}-C_{1})Q^{*}_{1}(x_{0}-x,t_{0}-t)\partial_{x}Q_{1}(x,t)\bigg]\,dx-i\Upsilon\,,\\
    \mathcal{C}_{3}^{(2)}&=&\int_{\mathbb{R}}\bigg[\epsilon_{1}(C_{3}-C_{1})Q^{*}_{1}(x_{0}-x,t_{0}-t)\partial_{x}Q_{1}(x,t)\nonumber\\&&\qquad-\,\epsilon_{2}(C_{3}-C_{2})Q^{*}_{2}(x_{0}-x,t_{0}-t)\partial_{x}Q_{2}(x,t)\bigg]\,dx-i\Upsilon\,,
\end{eqnarray}
\ese
where in the above, we have defined
\begin{equation}
\label{triple}
    \Upsilon=\int_{\mathbb{R}}\bigg[Q_{1}(x,t)Q_{2}(x,t)Q_{3}(x,t)-Q_{1}^{*}(x_{0}-x,t_{0}-t)Q_{2}^{*}(x_{0}-x,t_{0}-t)Q_{3}^{*}(x_{0}-x,t_{0}-t)\bigg]\,dx\,.
\end{equation}
Again, note that one of these quantities could be written as a linear combination of the other two.
%%%%%%%%%%%%%%%%%%%%%%%%%%%%%%%%%%%%%%%%%%%%%%%%%%%%%
%%%%%%%%%%%%%%%%%%%%%%%%%%%%%%%%%%%%%%%%%%%%%%%%%%%%%%%%%%%%%%%%%%%%
\subsection{Hamiltonian structure}
It turns out that the second set of conserved quantities \eqref{new_quantities} that one obtains from the infinite recurrence is directly connected to the infinite-dimensional Hamiltonian structure of the integrable six-wave interaction system. Particularly, a Hamiltonian for \eqref{six_wave_QR} is:
%%%%%%%%%%%%%%%%%%%%%%%%%%%%%%%%%%%%%%%%%%%%%%%%%%%%%%%%%%%%%%%%%
\begin{equation}
\label{hamiltonian}
\mathcal{H}=-i\int_{\mathbb{R}}\bigg[C_{1}R_{1}\partial_{x}Q_{1}+C_{2}R_{2}\partial_{x}Q_{2}+C_{3}R_{3}\partial_{x}Q_{3}+iQ_{1}Q_{2}Q_{3}+iR_{1}R_{2}R_{3}\bigg]\,dx\,.
\end{equation}
%%%%%%%%%%%%%%%%%%%%%%%%%%%%%%%%%%%%%%%%%%%%%%%%%%%%%%%%%%%%%%%%%
Indeed, one can verify that \eqref{six_wave_QR} is equivalent to 
\begin{equation}
\label{hamiltons_eq}
    i\partial_{t}Q_{j}=\frac{\delta\mathcal{H}}{\delta R_{j}}\,,\qquad i\partial_{t}R_{j}=-\frac{\delta\mathcal{H}}{\delta Q_{j}}\,.
\end{equation}
%%%%%%%%%%%%%%%%%%%%%%%%%%%%%%%%%%%%%%%%%%%%%%%%%%%%%%%%%%%%%%%%%
Furthermore, if we define the Poisson bracket
\begin{equation}
    \{F,G\}=\sum_{k=1}^{3}\int_{\mathbb{R}}\bigg(\frac{\delta F}{\delta Q_{k}(z)}\frac{\delta G}{\delta R_{k}(z)}-\frac{\delta F}{\delta R_{k}(z)}\frac{\delta G}{\delta Q_{k}(z)}\bigg)\,dz\,,
\end{equation}
then the canonical variables $Q_{j}(x)$ and $R_{j}(x)$ satisfy
%%%%%%%%%%%%%%%%%%%%%%%%%%%%%%%%%%%%%%%%%%%%%%%%%%%%%%%%%%%%%%%%%
\bse
\begin{eqnarray}
    &\{Q_{j}(x),R_{j'}(x')\}=\delta_{jj'}\delta(x-x')\,,\\
    &\{Q_{j}(x),Q_{j'}(x')\}=0\,,\;\;\;\{R_{j}(x),R_{j'}(x')\}=0\,.
\end{eqnarray}
\ese
We also have that
%%%%%%%%%%%%%%%%%%%%%%%%%%%%%%%%%%%%%%%%%%%%%%%%%%%%%%%%%%%%%%%%%
\begin{equation}
    \{F,Q_{j}(x)\}=-\frac{\delta F}{\delta R_{j}(x)}\,,\qquad\{F,R_{j}(x)\}=\frac{\delta F}{\delta Q_{j}(x)}\,,
\end{equation}
%%%%%%%%%%%%%%%%%%%%%%%%%%%%%%%%%%%%%%%%%%%%%%%%%%%%%%%%%%%%%%%%%
in which case Hamilton's equations \eqref{hamiltons_eq} can be expressed in terms of the bracket as
%%%%%%%%%%%%%%%%%%%%%%%%%%%%%%%%%%%%%%%%%%%%%%%%%%%%%%%%%%%%%%%%%
\begin{equation}
\label{hamilton_bracket}
    i\partial_{t}Q_{j}=\{Q_{j},\mathcal{H}\}\,,\qquad i\partial_{t}R_{j}=\{R_{j},\mathcal{H}\}\,.
\end{equation}
%%%%%%%%%%%%%%%%%%%%%%%%%%%%%%%%%%%%%%%%%%%%%%%%%%%%%%%%%%%%%%%%%
Observe that the Hamiltonian \eqref{hamiltonian} can be obtained as a linear combination of the three conserved quantities listed in \eqref{new_quantities} (in many ways, since they are linearly dependent). Specifically, 
%%%%%%%%%%%%%%%%%%%%%%%%%%%%%%%%%%%%%%%%%%%%%%%%%%%%%%%%%%%%%%%%%
\begin{equation}
    -i\mathcal{H}=\eta_{1}\mathcal{C}_{1}^{(2)}+\eta_{2}\mathcal{C}^{(2)}_{2}+\eta_{3}\mathcal{C}^{(2)}_{3}\,,
\end{equation}
%%%%%%%%%%%%%%%%%%%%%%%%%%%%%%%%%%%%%%%%%%%%%%%%%%%%%%%%%%%%%%%%%
for any $\eta_{1},\eta_{2},\eta_{3}$ satisfying the conditions 
$C_{1}\eta_{1}+C_{2}\eta_{2}+C_{3}\eta_{3}=0$ and $\eta_{1}+\eta_{2}+\eta_{3}=1$. Applying the symmetry reduction \eqref{reduc_classical}, the Hamiltonian for the classical three-wave system is
%%%%%%%%%%%%%%%%%%%%%%%%%%%%%%%%%%%%%%%%%%%%%%%%%%%%%%%%%%%%%%%%%
\begin{equation}
\mathcal{H}=i\int_{\mathbb{R}}\bigg[\tilde\epsilon_{1}C_{1}Q^{*}_{1}\partial_{x}Q_{1}+\tilde\epsilon_{2}C_{2}Q^{*}_{2}\partial_{x}Q_{2}+\tilde\epsilon_{3}C_{3}Q^{*}_{3}\partial_{x}Q_{3}\bigg]\,dx+2\,\text{Re}\int_{\mathbb{R}}Q_{1}Q_{2}Q_{3}\,dx\,.
\end{equation}
%%%%%%%%%%%%%%%%%%%%%%%%%%%%%%%%%%%%%%%%%%%%%%%%%%%%%%%%%%%%%%%%%
Note that using integration by parts, one can verify that $\mathcal{H}$ is real. Finally, applying \eqref{reduc_shifted}, the Hamiltonian for the space-time shifted three wave system is
%%%%%%%%%%%%%%%%%%%%%%%%%%%%%%%%%%%%%%%%%%%%%%%%%%%%%%%%%%%%%%%%%
\begin{eqnarray}
\mathcal{H}&=&-i\int_{\mathbb{R}}\bigg[\epsilon_{1}C_{1}Q^{*}_{1}(x_{0}-x,t_{0}-t)\partial_{x}Q_{1}(x,t)-\epsilon_{2}C_{2}Q^{*}_{2}(x_{0}-x,t_{0}-t)\partial_{x}Q_{2}(x,t)\nonumber\\&&\qquad+\,\epsilon_{3}C_{3}Q^{*}_{3}(x_{0}-x,t_{0}-t)\partial_{x}Q_{3}(x,t)\bigg]\,dx+\Upsilon\,,
\end{eqnarray}
%%%%%%%%%%%%%%%%%%%%%%%%%%%%%%%%%%%%%%%%%%%%%%%%%%%%%%%%%%%%%%%%%
with $\Upsilon$ defined as in \eqref{triple}. Using integration by parts, changing the variable of integration $x\rightarrow x_{0}-x$, and letting $t\rightarrow t_{0}-t$ (since $\mathcal{H}$ is conserved) shows that $\mathcal{H}$ is purely imaginary, as required by \eqref{hamiltons_eq} with the reduction \eqref{reduc_shifted}.
%%%%%%%%%%%%%%%%%%%%%%%%%%%%%%%%%%%%%%%%%%%%%%%%%%%%%%%%%%%%%%%%%%%%
\subsection{Conserved quantities and Hamiltonian structure in 2+1 dimensions}
%%%%%%%%%%%%%%%%%%%%%%%%%%%%%%%%%%%%%%%%%%%%%%%%%%%%%
We remark that one can generalize the above conservation laws to two spatial dimensions by inspection. The first set of conserved quantities \eqref{old_quantities} becomes:
\bse
\begin{eqnarray}
    \mathcal{C}_{1}^{(1)}&=&\iint_{\mathbb{R}^2}Q_{2}R_{2}\,dx\,dy - \iint_{\mathbb{R}^2}Q_{3}R_{3}\,dx\,dy\,,\\
    \mathcal{C}_{2}^{(1)}&=&\iint_{\mathbb{R}^2}Q_{3}R_{3}\,dx\,dy - \iint_{\mathbb{R}^2}Q_{1}R_{1}\,dx\,dy\,,\\
    \mathcal{C}_{3}^{(1)}&=&\iint_{\mathbb{R}^2}Q_{1}R_{1}\,dx\,dy - \iint_{\mathbb{R}^2}Q_{2}R_{2}\,dx\,dy\,.
\end{eqnarray}
\ese
%%%%%%%%%%%%%%%%%%%%%%%%%%%%%%%%%%%%%%%%%%%%%%%%%%%%%%
Incorporating the symmetry reduction \eqref{reduc_shifted}, in the space-time shifted case these are
%%%%%%%%%%%%%%%%%%%%%%%%%%%%%%%%%%%%%%%%%%%%%%%%%%%%%%%%%%%%%
\bse
\begin{eqnarray}
    \mathcal{C}_{1}^{(1)}&=&\epsilon_{2}\iint_{\mathbb{R}^{2}}Q_{2}(x,y,t)Q_{2}^{*}(x_{0}-x,y_{0}-y,t_{0}-t)\,dx\,dy\nonumber\\&&+\,\epsilon_{3}\iint_{\mathbb{R}^2}Q_{3}(x,y,t)Q_{3}^{*}(x_{0}-x,y_{0}-y,t_{0}-t)\,dx\,dy\,,\\
    \mathcal{C}_{2}^{(1)}&=&\epsilon_{3}\iint_{\mathbb{R}^2}Q_{3}(x,y,t)Q_{3}^{*}(x_{0}-x,y_{0}-y,t_{0}-t)\,dx\,dy\nonumber\\&&-\,\epsilon_{1}\iint_{\mathbb{R}^2}Q_{1}(x,y,t)Q_{1}^{*}(x_{0}-x,y_{0}-y,t_{0}-t)\,dx\,dy\,,\\
    \mathcal{C}_{3}^{(1)}&=&\epsilon_{1}\iint_{\mathbb{R}^2}Q_{1}(x,y,t)Q_{1}^{*}(x_{0}-x,y_{0}-y,t_{0}-t)\,dx\,dy\nonumber\\&&+\,\epsilon_{2}\iint_{\mathbb{R}^2}Q_{2}(x,y,t)Q_{2}^{*}(x_{0}-x,y_{0}-y,t_{0}-t)\,dx\,dy\,.
\end{eqnarray}
\ese
%%%%%%%%%%%%%%%%%%%%%%%%%%%%%%%%%%%%%%%%%%%%%%%%%%%%%%%%%%%%%
The generalization of the second set of conserved quantities \eqref{new_quantities} is given by:
%%%%%%%%%%%%%%%%%%%%%%%%%%%%%%%%%%%%%%%%%%%%%%%%%%%%%%%%%%%%%
\bse
\begin{eqnarray}
    \mathcal{C}_{1}^{(2)}&=&\iint_{\mathbb{R}^2}\bigg[(C^{(x)}_{1}-C^{(x)}_{2})R_{2}\partial_{x}Q_{2}+(C^{(x)}_{1}-C^{(x)}_{3})R_{3}\partial_{x}Q_{3}\nonumber\\&&+\,(C^{(y)}_{1}-C^{(y)}_{2})R_{2}\partial_{y}Q_{2}+(C^{(y)}_{1}-C^{(y)}_{3})R_{3}\partial_{y}Q_{3}-iQ_{1}Q_{2}Q_{3}-iR_{1}R_{2}R_{3}\bigg]\,dx\,dy\,,\\
    \mathcal{C}_{2}^{(2)}&=&\iint_{\mathbb{R}^2}\bigg[(C^{(x)}_{2}-C^{(x)}_{3})R_{3}\partial_{x}Q_{3}+(C^{(x)}_{2}-C^{(x)}_{1})R_{1}\partial_{x}Q_{1}\nonumber\\&&+\,(C^{(y)}_{2}-C^{(y)}_{3})R_{3}\partial_{y}Q_{3}+(C^{(y)}_{2}-C^{(y)}_{1})R_{1}\partial_{y}Q_{1}-iQ_{1}Q_{2}Q_{3}-iR_{1}R_{2}R_{3}\bigg]\,dx\,dy\,,\\
    \mathcal{C}_{3}^{(2)}&=&\iint_{\mathbb{R}^2}\bigg[(C^{(x)}_{3}-C^{(x)}_{1})R_{1}\partial_{x}Q_{1}+(C^{(x)}_{3}-C^{(x)}_{2})R_{2}\partial_{x}Q_{2}\nonumber\\&&+\,(C^{(y)}_{3}-C^{(y)}_{1})R_{1}\partial_{y}Q_{1}+(C^{(y)}_{3}-C^{(y)}_{2})R_{2}\partial_{y}Q_{2}-iQ_{1}Q_{2}Q_{3}-iR_{1}R_{2}R_{3}\bigg]\,dx\,dy\,.
\end{eqnarray}
\ese
%%%%%%%%%%%%%%%%%%%%%%%%%%%%%%%%%%%%%%%%%%%%%%%%%%%%%%%%%%%%%%%%%
Applying the shifted symmetry reduction \eqref{reduc_shifted} gives
%%%%%%%%%%%%%%%%%%%%%%%%%%%%%%%%%%%%%%%%%%%%%%%%%%%%%%%%%%%%%%%%%
\bse
\begin{eqnarray}
    \mathcal{C}_{1}^{(2)}&=&\iint_{\mathbb{R}^2}\bigg[-\epsilon_{2}(C_{1}^{(x)}-C^{(x)}_{2})Q^{*}_{2}(x_{0}-x,y_{0}-y,t_{0}-t)\partial_{x}Q_{2}(x,y,t)\nonumber\\&&\qquad\qquad+\,\epsilon_{3}(C^{(x)}_{1}-C^{(x)}_{3})Q^{*}_{3}(x_{0}-x,y_{0}-y,t_{0}-t)\partial_{x}Q_{3}(x,y,t)\nonumber\\&&\qquad\qquad-\,\epsilon_{2}(C_{1}^{(y)}-C^{(y)}_{2})Q^{*}_{2}(x_{0}-x,y_{0}-y,t_{0}-t)\partial_{y}Q_{2}(x,y,t)\nonumber\\&&\qquad\qquad+\,\epsilon_{3}(C^{(y)}_{1}-C^{(y)}_{3})Q^{*}_{3}(x_{0}-x,y_{0}-y,t_{0}-t)\partial_{y}Q_{3}(x,y,t)\bigg]\,dx\,dy-i\Upsilon\,,
    \end{eqnarray}
    %%%%%%%%%%%%%%%%%%%%%%%%%%%%%%%%%%%%%%%%%%%%%%%%%%%%%%%%%%%%%%%%%
    \begin{eqnarray}
    %%%%%%%%%%%%%%%%%%%%%%%%%%%%%%%%%%%%%%%%%%%%%%%%%%%%%%%%%%%%%%%%%
    \mathcal{C}_{2}^{(2)}&=&\iint_{\mathbb{R}^2}\bigg[\epsilon_{3}(C^{(x)}_{2}-C^{(x)}_{3})Q^{*}_{3}(x_{0}-x,y_{0}-y,t_{0}-t)\partial_{x}Q_{3}(x,y,t)\nonumber\\&&\qquad\qquad+\,\epsilon_{1}(C^{(x)}_{2}-C^{(x)}_{1})Q^{*}_{1}(x_{0}-x,y_{0}-y,t_{0}-t)\partial_{x}Q_{1}(x,y,t)\nonumber\\&&\qquad\qquad+\,\epsilon_{3}(C^{(y)}_{2}-C^{(y)}_{3})Q^{*}_{3}(x_{0}-x,y_{0}-y,t_{0}-t)\partial_{y}Q_{3}(x,y,t)\nonumber\\&&\qquad\qquad+\,\epsilon_{1}(C^{(y)}_{2}-C^{(y)}_{1})Q^{*}_{1}(x_{0}-x,y_{0}-y,t_{0}-t)\partial_{y}Q_{1}(x,y,t)\bigg]\,dx\,dy-i\Upsilon\,,
    \end{eqnarray}
    %%%%%%%%%%%%%%%%%%%%%%%%%%%%%%%%%%%%%%%%%%%%%%%%%%%%%%%%%%%%%%%%%
    \begin{eqnarray}
    %%%%%%%%%%%%%%%%%%%%%%%%%%%%%%%%%%%%%%%%%%%%%%%%%%%%%%%%%%%%%%%%%
    \mathcal{C}_{3}^{(2)}&=&\iint_{\mathbb{R}^2}\bigg[\epsilon_{1}(C^{(x)}_{3}-C^{(x)}_{1})Q^{*}_{1}(x_{0}-x,y_{0}-y,t_{0}-t)\partial_{x}Q_{1}(x,y,t)\nonumber\\&&\qquad\qquad-\,\epsilon_{2}(C^{(x)}_{3}-C^{(x)}_{2})Q^{*}_{2}(x_{0}-x,y_{0}-y,t_{0}-t)\partial_{x}Q_{2}(x,y,t)\nonumber\\&&\qquad\qquad+\,\epsilon_{1}(C^{(y)}_{3}-C^{(y)}_{1})Q^{*}_{1}(x_{0}-x,y_{0}-y,t_{0}-t)\partial_{y}Q_{1}(x,y,t)\nonumber\\&&\qquad\qquad-\,\epsilon_{2}(C^{(y)}_{3}-C^{(y)}_{2})Q^{*}_{2}(x_{0}-x,y_{0}-y,t_{0}-t)\partial_{y}Q_{2}(x,y,t)\bigg]\,dx\,dy-i\Upsilon\,,
\end{eqnarray}
\ese
%%%%%%%%%%%%%%%%%%%%%%%%%%%%%%%%%%%%%%%%%%%%%%%%%%%%%%%%%%%%%%%%%
where 
%%%%%%%%%%%%%%%%%%%%%%%%%%%%%%%%%%%%%%%%%%%%%%%%%%%%%%%%%%%%%%%%%
\begin{eqnarray}
    \Upsilon&=&\iint_{\mathbb{R}}\bigg[Q_{1}(x,y,t)Q_{2}(x,y,t)Q_{3}(x,y,t)\nonumber\\&&-\,Q_{1}^{*}(x_{0}-x,y_{0}-y,t_{0}-t)Q_{2}^{*}(x_{0}-x,y_{0}-y,t_{0}-t)Q_{3}^{*}(x_{0}-x,y_{0}-y,t_{0}-t)\bigg]\,dx\,dy\,.
\end{eqnarray}
%%%%%%%%%%%%%%%%%%%%%%%%%%%%%%%%%%%%%%%%%%%%%%%%%%%%%%%%%%%%%%%%%
Finally, the Hamiltonian \eqref{hamiltonian} for the $1+1$ dimensional six-wave system can also be generalized to the $2+1$ dimensional six-wave system:
%%%%%%%%%%%%%%%%%%%%%%%%%%%%%%%%%%%%%%%%%%%%%%%%%%%%%%%%%%%%%%%%%
\begin{eqnarray}
\mathcal{H}&=&-i\iint_{\mathbb{R}^2}\bigg[C^{(x)}_{1}R_{1}\partial_{x}Q_{1}+C^{(x)}_{2}R_{2}\partial_{x}Q_{2}+C^{(x)}_{3}R_{3}\partial_{x}Q_{3}\nonumber\\&&\qquad+\,C^{(y)}_{1}R_{1}\partial_{y}Q_{1}+C^{(y)}_{2}R_{2}\partial_{y}Q_{2}+C^{(y)}_{3}R_{3}\partial_{y}Q_{3}+iQ_{1}Q_{2}Q_{3}+iR_{1}R_{2}R_{3}\bigg]\,dx\,dy\,.
\end{eqnarray}
%%%%%%%%%%%%%%%%%%%%%%%%%%%%%%%%%%%%%%%%%%%%%%%%%%%%%%%%%%%%%%%%%
The $(2+1)$ dimensional version of the six-wave system is then equivalent to \eqref{hamiltons_eq} or \eqref{hamilton_bracket} with the Poisson bracket now defined by
%%%%%%%%%%%%%%%%%%%%%%%%%%%%%%%%%%%%%%%%%%%%%%%%%%%%%%%%%%%%%%%%%
\begin{equation}
    \{F,G\}=\sum_{k=1}^{3}\iint_{\mathbb{R}^2}\bigg(\frac{\delta F}{\delta Q_{k}(z_{1},z_{2})}\frac{\delta G}{\delta R_{k}(z_{1},z_{2})}-\frac{\delta F}{\delta R_{k}(z_{1},z_{2})}\frac{\delta G}{\delta Q_{k}(z_{1},z_{2})}\bigg)\,dz_{1}\,dz_{2}\,.
\end{equation}
%%%%%%%%%%%%%%%%%%%%%%%%%%%%%%%%%%%%%%%%%%%%%%%%%%%%%%%%%%%%%%%%%
In particular, under the space-time shifted reduction, the $2+1$ dimensional space-time shifted Hamiltonian is given by
%%%%%%%%%%%%%%%%%%%%%%%%%%%%%%%%%%%%%%%%%%%%%%%%%%%%%%%%%%%%%%%%%
\begin{eqnarray}
\mathcal{H}&=&-i\iint_{\mathbb{R}^2}\bigg[\epsilon_{1}C^{(x)}_{1}Q^{*}_{1}(x_{0}-x,y_{0}-y,t_{0}-t)\partial_{x}Q_{1}(x,y,t)\nonumber\\&&\qquad\qquad-\,\epsilon_{2}C^{(x)}_{2}Q^{*}_{2}(x_{0}-x,y_{0}-y,t_{0}-t)\partial_{x}Q_{2}(x,y,t)\nonumber\\&&\qquad\qquad+\,\epsilon_{3}C^{(x)}_{3}Q^{*}_{3}(x_{0}-x,y_{0}-y,t_{0}-t)\partial_{x}Q_{3}(x,y,t)\nonumber\\&&\qquad\qquad+\,\epsilon_{1}C^{(y)}_{1}Q^{*}_{1}(x_{0}-x,y_{0}-y,t_{0}-t)\partial_{y}Q_{1}(x,y,t)\nonumber\\&&\qquad\qquad-\,\epsilon_{2}C^{(y)}_{2}Q^{*}_{2}(x_{0}-x,y_{0}-y,t_{0}-t)\partial_{y}Q_{2}(x,y,t)\nonumber\\&&\qquad\qquad+\,\epsilon_{3}C^{(y)}_{3}Q^{*}_{3}(x_{0}-x,y_{0}-y,t_{0}-t)\partial_{y}Q_{3}(x,y,t)\bigg]\,dx\,dy+\Upsilon\,.
\end{eqnarray}
%%%%%%%%%%%%%%%%%%%%%%%%%%%%%%%%%%%%%%%%%%%%%%%%%%%%%%%%%%%%%%%%%
%%%%%%%%%%%%%%%%%%%%%%%%%%%%%%%%%%%%%%%%%%%%%%%%%%%%%%%%%%%%%%%%%
\section{Conclusion}
%%%%%%%%%%%%%%%%%%%%%%%%%%%%%%%%%%%%%%%%%%%%%%%%%%%%%%%%%%%%%%%%%%
%%%%%%%%%%%%%%%%%%%%%%%%%%%%%%%%%%%%%%%%%%%%%%%%%%%%%%%%%%%%%%%%%%%%%	
In this paper, $(2+1)$ dimensional first order quadratically coupled nonlinear six-wave interaction equations were derived from the physics of 
nonlinear optics. This was accomplished by considering two model examples. The first is based on Maxwell's equations where the polarization field 
is expressed in terms of a time-convolution between the electric field and medium susceptibility. The other case is based on a laser model where the polarization function satisfies a quadratically nonlinear differential equation. Upon employing a space-time multi-scale 
asymptotic expansion (on both models), a hierarchy of linearly coupled nonhomogeneous equations is derived. Importantly, the leading order equation admits a solution in the form of a linear superposition of six plane waves whose wave-vector and corresponding frequency satisfy the triad resonance condition 
(\ref{resonance}). Importantly, the slowly modulated amplitudes of each individual plane wave are not assumed to be complex conjugates of each other. 
By removing secular terms at order $\epsilon$ (the small expansion parameter), first order in space and time quadratically coupled six-wave 
equations in $2+1$ dimensions are obtained. 
Remarkably, this resulting system is shown to be connected to its integrable counterpart which was mathematically derived by Ablowitz and Haberman in the 1970s. Several integrable reductions to space-time shifted nonlocal three-wave equations are obtained and shown to form an integrable Hamiltonian system. In this regard, a set of integrals of motions are derived. The nonlocal space-time shifted three-wave equations are solved using the inverse scattering transform. Nonlocal symmetries between the associated eigenfunctions and scattering data are derived. 
A Riemann-Hilbert problem is formulated which is used to obtain soliton solutions.
%%%%%%%%%%%%%%%%%%%%%%%%%%%%%%%%%%%%%%%%%%%%%%%%%%%%%%%%%%%%%%%%%%

\section*{Acknowledgements}
MJA was partially supported by NSF under Grant No. DMS-2306290.

\appendix
\section*{Appendix}
%%%%%%%%%%%%%%%%%%%%%%%%%%%%%%%%%%%%%%%%%%%%%%%%%%%%%%%%%%%%%%%%%%
In this Appendix, we list all nonlinear terms that contribute to a resonant three wave triad and show how (3.40) arise 
from (\ref{six-waves-simple}). We shall assume that the nonlinear susceptibility satisfies the two symmetry conditions: (i) 
$\hat{\chi}_{zzz}^{(NL)}(-\zeta_1,-\zeta_2) = \hat{\chi}_{zzz}^{(NL)}(\zeta_1,\zeta_2)$ and (ii)
$\hat{\chi}_{zzz}^{(NL)}(\zeta_1,\zeta_2) = \hat{\chi}_{zzz}^{(NL)}(\zeta_2,\zeta_1)$.
Note that the only nonlinear terms in (\ref{six-waves-simple}) that lead to resonance are the follows.\\\\ 
%%%%%%%%%%%%%%%%%%%%%%%%%%%%%%%%%%%%%%%%%%%%%%%%%%%%%%%%%%%%%%%%%%%%%
%%%%%%%%%%%%%%%%%%%%%%%%%%%%%%%%%%%%%%%%%%%%%%%%%%%%%%%%%%%%%%%%%%%%%
Terms proportional to $e^{i\theta_1}:$
%%%%%%%%%%%%%%%%%%%%
\begin{itemize}
%%%%%%%%%%%%%%%%%%%%
\item $n=2, m=3\;, \;\;\;\;\; 
(\omega_2 + \omega_3 )^2 \hat{\chi}_{zzz}^{(NL)}(-\omega_2,-\omega_3) 
     B_{2} B_{3}$
%%%%%%%%%%%%%%%%%%%%
\item $n=3, m=2\;, \;\;\;\;\; 
(\omega_3 + \omega_2 )^2 \hat{\chi}_{zzz}^{(NL)}(-\omega_3,-\omega_2) 
     B_{3} B_{2}$
\end{itemize}
%%%%%%%%%%%%%%%%%%%%%%%%%%%%%%%%%%%%%%%%%%%%%%%%%%%%%%%%%%%%%%%%%%%%%%%%%%%%%
%%%%%%%%%%%%%%%%%%%%%%%%%%%%%%%%%%%%%%%%%%%%%%%%%%%%%%%%%%%%%%%%%%%%%
Terms proportional to $e^{i\theta_2}:$
%%%%%%%%%%%%%%%%%%%%
\begin{itemize}
%%%%%%%%%%%%%%%%%%%%
\item $n=1, m=3\;, \;\;\;\;\; 
(\omega_1 + \omega_3 )^2  \hat{\chi}_{zzz}^{(NL)}(-\omega_1,-\omega_3) 
     B_{1} B_{3}  $
%%%%%%%%%%%%%%%%%%%%
\item $n=3, m=1\;, \;\;\;\;\; 
 (\omega_3 + \omega_1 )^2  \hat{\chi}_{zzz}^{(NL)}(-\omega_3,-\omega_1) 
     B_{3} B_{1} $
\end{itemize}
%%%%%%%%%%%%%%%%%%%%%%%%%%%%%%%%%%%%%%%%%%%%%%%%%%%%%%%%%%%%%%%%%%%%%%%%%%%%%
%%%%%%%%%%%%%%%%%%%%%%%%%%%%%%%%%%%%%%%%%%%%%%%%%%%%%%%%%%%%%%%%%%%%%
Terms proportional to $e^{i\theta_3}:$
%%%%%%%%%%%%%%%%%%%%
\begin{itemize}
%%%%%%%%%%%%%%%%%%%%
\item $n=1, m=2\;, \;\;\;\;\; 
 (\omega_1 + \omega_2 )^2  \hat{\chi}_{zzz}^{(NL)}(-\omega_1,-\omega_2) 
     B_{1} B_{2}$
%%%%%%%%%%%%%%%%%%%%
\item $n=2, m=1\;, \;\;\;\;\; 
 (\omega_2 + \omega_1 )^2 \hat{\chi}_{zzz}^{(NL)}(-\omega_2,-\omega_1) 
     B_{2} B_{1}$
\end{itemize}
%%%%%%%%%%%%%%%%%%%%%%%%%%%%%%%%%%%%%%%%%%%%%%%%%%%%%%%%%%%%%%%%%%%%%%%%%
%%%%%%%%%%%%%%%%%%%%%%%%%%%%%%%%%%%%%%%%%%%%%%%%%%%%%%%%%%%%%%%%%%%%%
Terms proportional to $e^{-i\theta_1}:$
%%%%%%%%%%%%%%%%%%%%
\begin{itemize}
%%%%%%%%%%%%%%%%%%%%
\item $n=2, m=3\;, \;\;\;\;\; 
  (\omega_2 + \omega_3 )^2 \hat{\chi}_{zzz}^{(NL)}(\omega_2,\omega_3) 
     A_{2} A_{3}$
%%%%%%%%%%%%%%%%%%%%
\item $n=3, m=2\;, \;\;\;\;\; 
 (\omega_3 + \omega_2 )^2 \hat{\chi}_{zzz}^{(NL)}(\omega_3,\omega_2) 
     A_{3} A_{2} $
\end{itemize}
%%%%%%%%%%%%%%%%%%%%%%%%%%%%%%%%%%%%%%%%%%%%%%%%%%%%%%%%%%%%%%%%%%%%%%%%%%%%%
%%%%%%%%%%%%%%%%%%%%%%%%%%%%%%%%%%%%%%%%%%%%%%%%%%%%%%%%%%%%%%%%%%%%%
Terms proportional to $e^{-i\theta_2}:$
%%%%%%%%%%%%%%%%%%%%
\begin{itemize}
%%%%%%%%%%%%%%%%%%%%
\item $n=1, m=3\;, \;\;\;\;\; 
  (\omega_1 + \omega_3 )^2  \hat{\chi}_{zzz}^{(NL)}(\omega_1,\omega_3) 
     A_{1} A_{3}$
%%%%%%%%%%%%%%%%%%%%
\item $n=3, m=1\;, \;\;\;\;\; 
  (\omega_3 + \omega_1 )^2 \hat{\chi}_{zzz}^{(NL)}(\omega_3,\omega_1) 
     A_{3} A_{1} $
\end{itemize}
%%%%%%%%%%%%%%%%%%%%%%%%%%%%%%%%%%%%%%%%%%%%%%%%%%%%%%%%%%%%%%%%%%%%%%%%%%%
%%%%%%%%%%%%%%%%%%%%%%%%%%%%%%%%%%%%%%%%%%%%%%%%%%%%%%%%%%%%%%%%%%%%%
Terms proportional to $e^{-i\theta_3}:$
%%%%%%%%%%%%%%%%%%%%
\begin{itemize}
%%%%%%%%%%%%%%%%%%%%
\item $n=1, m=2\;, \;\;\;\;\; 
 (\omega_1 + \omega_2 )^2\hat{\chi}_{zzz}^{(NL)}(\omega_1,\omega_2) 
     A_{1} A_{2} $
%%%%%%%%%%%%%%%%%%%%
\item $n=2, m=1\;, \;\;\;\;\; 
  (\omega_2 + \omega_1 )^2 \hat{\chi}_{zzz}^{(NL)}(\omega_2,\omega_1) 
     A_{2} A_{1}$
\end{itemize}
%%%%%%%%%%%%%%%%%%%%%%%%%%%%%%%%%%%%%%%%%%%%%%%%%%%%%%%%%%%%%%%%%%%%%%%%%%%%
With this at hand, we now collect all (linear and nonlinear) terms that contribute to resonance triad which gives rise to the six wave resonance equations.\\\\
%%%%%%%%%%%%%%%%%%%%%%%%%%%%%%%%%%%%%%%%%%%%%%%%%%%%%%%%%%%%%%%%%%%%%%%%%%%
Coefficient of  $e^{i\theta_1}:$
%%%%%%%%%%%%%%%%%%%%
\begin{align*}
 2i\omega_1 \left( 1 +  \hat{\chi}_{zz}^{(L)}(\omega_1) 
  +  \frac{\omega_1}{2} \partial_\omega \hat{\chi}_{zz}^{(L)}(\omega_1)  \right) \frac{\partial A_1}{\partial T} 
     +  2i c^2 (\mathbf{k}_1  \cdot \nabla_{\mathbf{R}_{\perp}} ) A_1
%\nonumber \\
+
2\omega^2_1 \hat{\chi}_{zzz}^{(NL)}(\omega_2,\omega_3)  B_{3} B_{2}
   =0 \;.
 \end{align*}
%%%%%%%%%%%%%%%%%%%%%%%%%%%%%%%%%%%%%%%%%%%%%%%%%%%%%%%%%%%%%%%%%%%%%
%%%%%%%%%%%%%%%%%%%%%%%%%%%%%%%%%%%%%%%%%%%%%%%%%%%%%%%%%%%%%%%%%%%
Coefficient of  $e^{i\theta_2}:$
%%%%%%%%%%%%%%%%%%%%
\begin{align*}
 2i\omega_2 \left( 1 +  \hat{\chi}_{zz}^{(L)}(\omega_2) 
  +  \frac{\omega_1}{2} \partial_\omega \hat{\chi}_{zz}^{(L)}(\omega_2)  \right) \frac{\partial A_2}{\partial T} 
     +  2i c^2 (\mathbf{k}_2  \cdot \nabla_{\mathbf{R}_{\perp}} ) A_2
+
2 \omega^2_2   \hat{\chi}_{zzz}^{(NL)}(\omega_1,\omega_3) B_{1} B_{3}
   =0 \;.
 \end{align*}
%%%%%%%%%%%%%%%%%%%%%%%%%%%%%%%%%%%%%%%%%%%%%%%%%%%%%%%%%%%%%%%
%%%%%%%%%%%%%%%%%%%%%%%%%%%%%%%%%%%%%%%%%%%%%%%%%%%%%%%%%%%%%%%%%%%
Coefficient of  $e^{i\theta_3}:$
%%%%%%%%%%%%%%%%%%%%
\begin{align*}
 2i\omega_3 \left( 1 +  \hat{\chi}_{zz}^{(L)}(\omega_3) 
  +  \frac{\omega_3}{2} \partial_\omega \hat{\chi}_{zz}^{(L)}(\omega_3)  \right) \frac{\partial A_3}{\partial T} 
     +  2i c^2(\mathbf{k}_3  \cdot \nabla_{\mathbf{R}_{\perp}} ) A_3
+ 
2 \omega^2_3 \hat{\chi}_{zzz}^{(NL)}(\omega_1,\omega_2)  B_{1} B_{2} 
   =0 \;.
 \end{align*}
%%%%%%%%%%%%%%%%%%%%%%%%%%%%%%%%%%%%%%%%%%%%%%%%%%%%%%%%%%%%%%%
%%%%%%%%%%%%%%%%%%%%%%%%%%%%%%%%%%%%%%%%%%%%%%%%%%%%%%%%%%%%%%%%%%%%%%
%%%%%%%%%%%%%%%%%%%%%%%%%%%%%%%%%%%%%%%%%%%%%%%%%%%%%%%%%%%%%%%%%%%%%%%%%%%
Coefficient of  $e^{-i\theta_1}:$
%%%%%%%%%%%%%%%%%%%%
\begin{align*}
 2i\omega_1 \left( 1 +  \hat{\chi}_{zz}^{(L)}(\omega_1) 
  +  \frac{\omega_1}{2} \partial_\omega \hat{\chi}_{zz}^{(L)}(\omega_1)  \right) \frac{\partial B_1}{\partial T} 
     +  2i c^2(\mathbf{k}_1  \cdot \nabla_{\mathbf{R}_{\perp}} ) B_1
- 
2 \omega^2_1  \hat{\chi}_{zzz}^{(NL)}(\omega_2,\omega_3) A_{3} A_{2}
   =0 \;.
 \end{align*}
%%%%%%%%%%%%%%%%%%%%%%%%%%%%%%%%%%%%%%%%%%%%%%%%%%%%%%%%%%%%%%%%%%%%%
%%%%%%%%%%%%%%%%%%%%%%%%%%%%%%%%%%%%%%%%%%%%%%%%%%%%%%%%%%%%%%%%%%%
Coefficient of  $e^{-i\theta_2}:$
%%%%%%%%%%%%%%%%%%%%
\begin{align*}
 2i\omega_2 \left( 1 +  \hat{\chi}_{zz}^{(L)}(\omega_2) 
  +  \frac{\omega_1}{2} \partial_\omega \hat{\chi}_{zz}^{(L)}(\omega_2)  \right) \frac{\partial B_2}{\partial T} 
     +  2i c^2 (\mathbf{k}_2  \cdot \nabla_{\mathbf{R}_{\perp}} ) B_2
- 
 2 \omega^2_2 \hat{\chi}_{zzz}^{(NL)}(\omega_1,\omega_3)  A_{1} A_{3}
   =0 \;.
 \end{align*}
%%%%%%%%%%%%%%%%%%%%%%%%%%%%%%%%%%%%%%%%%%%%%%%%%%%%%%%%%%%%%%%
%%%%%%%%%%%%%%%%%%%%%%%%%%%%%%%%%%%%%%%%%%%%%%%%%%%%%%%%%%%%%%%%%%%
Coefficient of  $e^{-i\theta_3}:$
%%%%%%%%%%%%%%%%%%%%
\begin{align*}
2i\omega_3 \left( 1 +  \hat{\chi}_{zz}^{(L)}(\omega_3) 
  +  \frac{\omega_3}{2} \partial_\omega \hat{\chi}_{zz}^{(L)}(\omega_3)  \right) \frac{\partial B_3}{\partial T} 
     +  2i c^2 (\mathbf{k}_3  \cdot \nabla_{\mathbf{R}_{\perp}} ) B_3
- 
2 \omega^2_3   \hat{\chi}_{zzz}^{(NL)}(\omega_1,\omega_2) A_{1} A_{2} 
   =0 \;.
 \end{align*}
%%%%%%%%%%%%%%%%%%%%%%%%%%%%%%%%%%%%%%%%%%%%%%%%%%%%%%%%%%%%%%%
%%%%%%%%%%%%%%%%%%%%%%%%%%%%%%%%%%%%%%%%%%%%%%%%%%%%%%%%%%%%%%%%%%%%%%
%%%%%%%%%%%%%%%%%%%%%%%%%%%%%%%%%%%%%%%%%%%%%%%%%%%%%%%%%%%%%%%%%%

\end{document}